\documentclass[useAMS,usenatbib,letterpaper]{mn2e}
\usepackage{graphicx,amsmath,color,amssymb}
\usepackage[pdftitle={Damped Lyman-alpha absorbers as a probe of feedback},pdfauthor={Simeon Bird}]{hyperref}
\usepackage[all]{hypcap}
\usepackage{subfigure}

\newlength{\narrowFigurewidth}
\newlength{\Figurewidth}
\newlength{\wideFigurewidth}
\newcommand{\doi}[1]{}

\newcommand{\etal}
  {et al.}

\newcommand{\Lya}{Lyman-$\alpha\;$}
\newcommand{\Msun}{\, h^{-1} M_\odot}
\newcommand{\Zsun}{Z_\odot}
\newcommand{\NHunit}{cm$^{-2}$}

\newcommand{\Mpch}{\, h^{-1} \mathrm{Mpc}}
\newcommand{\kpch}{\, h^{-1}\mathrm{kpc}}
\newcommand{\hMpc}{\, h \mathrm{Mpc}^{-1}}
\newcommand{\gadget}{{\small GADGET\,}}

\newcommand{\arepo}{{\small AREPO\,}}

\newcommand{\NHI}{N_\mathrm{HI}}
\newcommand{\kms}{km\,s$^{-1}$ }

\newcommand{\sigmaDLA}{\sigma_\mathrm{DLA}}

\title[DLAs as a probe of feedback]{Damped Lyman-$\alpha$ absorbers as a probe of stellar feedback}

\author
  [S. Bird \etal]
  {Simeon Bird$^{1}$\thanks{E-mail: spb@ias.edu},
  Mark Vogelsberger$^{2}$, Martin Haehnelt$^{3}$,  Debora Sijacki$^{3}$, Shy Genel$^{4}$, 
  \newauthor Paul Torrey$^{4}$, Volker Springel$^{5,6}$, Lars Hernquist$^{4}$ \vspace{4mm}\\
$^1$Institute for Advanced Study, 1 Einstein Drive, Princeton, NJ, 08540, USA\\
$^2$MKI and Department of Physics, Massachusetts Institute of Technology, 77 Massachusetts Avenue, Cambridge, MA 02139, USA \\
$^3$Kavli Institute for Cosmology and Institute for Astronomy, Madingley Road, Cambridge, CB3 0HA, UK \\
$^4$Harvard-Smithsonian Center for Astrophysics, 60 Garden Street, Cambridge, MA 02138, USA \\
$^5$Heidelberg Institute for Theoretical Studies, Schloss-Wolfsbrunnenweg 35, 69118 Heidelberg, Germany \\
$^6$Zentrum f\"{u}r Astronomie der Universit\"{a}t Heidelberg, ARI, M\"{o}nchhofstr. 12-14, 69120 Heidelberg, Germany \\
}

\begin{document}

\pagenumbering{alph}
\date{}

\maketitle
\pagerange{\protect\pageref{firstpage}--\protect\pageref{lastpage}} \pubyear{2012}

\pagenumbering{arabic}
\label{firstpage}

\begin{abstract}
We examine the abundance, clustering and metallicity of Damped Lyman-$\alpha$ Absorbers (DLAs) in a suite of hydrodynamic 
cosmological simulations using the moving mesh code \arepo. We incorporate models of supernova and AGN feedback, 
as well as molecular hydrogen formation. We compare our simulations to the column density distribution function at $z=3$, the 
total DLA abundance at $z=2-4$, the measured DLA bias at $z=2.3$ and the DLA metallicity distribution at $z=2-4$.
Our preferred models produce populations of DLAs in good agreement with most of these observations. The exception is the 
DLA abundance at $z < 3$, which we show requires stronger feedback in $10^{11-12} \Msun$ mass halos.
While the DLA population probes a wide range of halo masses, we find the cross-section is dominated by halos
of mass $10^{10} - 10^{11} \Msun$ and virial velocities $50 - 100$ \kms. The simulated DLA population has a 
linear theory bias of $1.7$, whereas the observations require $2.17 \pm 0.2$. We show that non-linear growth 
increases the bias in our simulations to $2.3$ at $k=1 \hMpc$, the smallest scale observed. 
The scale-dependence of the bias is, however, very different in the simulations compared against the observations.
We show that, of the observations we consider, the DLA abundance and column density function 
provide the strongest constraints on the feedback model.
\end{abstract}

 
\section{Introduction}
\label{sec:intro}

Damped \Lya Systems (DLAs) are strong \Lya~absorption features in quasar spectra.
At the high densities necessary to produce neutral hydrogen column densities 
above the threshold for DLAs, $\NHI > 10^{20.3}$ \NHunit \citep{Wolfe:1986}, hydrogen is 
self-shielded from the ionising effect of the diffuse radiation
background and so is neutral \citep{Katz:1996a}.
Because lower redshifts can only be observed from space, the bulk of observed
DLAs lie at $z=2-4$, where they contain a large fraction of the neutral hydrogen 
in the Universe \citep{Gardner:1997}. 
DLAs are a direct probe of the distribution of neutral gas 
at a mean density around one tenth of the star formation threshold \citep{Cen:2012}. 
Thus, they provide a powerful independent check on models of galaxy formation, with the unique 
advantage that they do not directly depend on the star formation model.
The exact nature of the systems hosting DLAs has historically been uncertain, with 
kinematic data combined with simple semi-analytic models appearing to favour large rotating discs 
\citep{Prochaska:1997, Jedamzik:1998, Maller:2001},
while early simulations produced small proto-galactic clumps \citep{Haehnelt:1998, Okoshi:2005}.
However, \cite{Barnes:2009} showed that the kinematic data could be reproduced if DLAs are found in halos 
with a virial velocity of $50-200$ \kms.

We have performed a series of cosmological hydrodynamic simulations with the 
moving mesh code \arepo\ \citep{Springel:2010}. \arepo~performs well in idealised tests \citep{Sijacki:2012} and yields galaxies with disk-like
morphologies \citep{Torrey:2012}. Compared to its predecessor \gadget, it follows the accretion of 
cold gas onto the central regions of halos significantly more accurately \citep{Nelson:2013}
and leads to a smoother distribution of Lyman Limit Systems (LLS) \citep{Bird:2013}.
We incorporate prescriptions for supernova and AGN feedback,
based on the implementation presented in \cite{Vogelsberger:2013} (V13) and \cite{Torrey:2013}.
Our simulations have been post-processed to produce a simulated DLA population, which we
compare to observations of the column density function, DLA metallicity, redshift evolution and bias.
These observations strongly constrain the stellar feedback model, and confirm the need for strong stellar 
feedback even at high redshifts. 

A wide range of quasar surveys have over time increased the available sample of high-redshift DLAs
\citep[e.g.][]{Wolfe:1995, Storrie:2000, Peroux:2005, Prochaska:2005, Noterdaeme:2012}, 
and constrained the distribution of systems at lower column densities
\citep{Peroux:2001,OMeara:2007, Prochaska:2009, Rudie:2013, Kim:2013, Zafar:2013}. 
Previous simulation work in this subject is also extensive.
\cite{Pontzen:2008} reproduced the observed DLA metallicity and abundance in SPH simulations
with a model incorporating supernova feedback and simple radiative transfer. 
The effect of galactic winds on DLA abundance and metallicity in SPH simulations was also examined in 
\cite{Nagamine:2004a, Nagamine:2004b, Nagamine:2007,Tescari:2009}.
Similar studies have been done with grid-based codes \citep{Fumagalli:2011, Cen:2012}.
\cite{FaucherGiguere:2010, Altay:2011, Yajima:2011, Rahmati:2013a} studied the effects of radiative transfer 
on DLA self-shielding. Molecular hydrogen formation was considered in \cite{Erkal:2012} and \cite{Altay:2011}.
\cite{Dave:2013} used a similar feedback scheme to this work, and examined the galactic HI mass function. 
\cite{Altay:2013} considered the column density distribution within
the models examined by the OWLS project, while \cite{Rahmati:2013c} looked at the halo hosts of 
both DLAs and Lyman Limit Systems (LLSs). \cite{Razoumov:2009} examined the kinematic distribution of DLA metal lines in
AMR simulations of isolated halos. \cite{Cen:2012} presented simulations of one over-dense and 
one under-dense region, which bracketed many observable properties of DLAs, 
including the kinematic distribution of metal lines, the metallicity and the column density function. 

In this paper, we compare DLAs from a hydrodynamic simulation to recently released data for the 
column density distribution and DLA abundance \citep{Noterdaeme:2012}, DLA metallicity \citep{Rafelski:2012} and 
the bias of DLAs \citep{FontRibera:2012}.
We use a suite of models based around that of V13, both providing a valuable consistency check on the model and 
demonstrating the constraints from each observation.
This paper is structured as follows. In Section~\ref{sec:methods}, we
discuss our methods. Section \ref{sec:properties} looks at the 
properties of our simulated DLAs, and Section \ref{sec:results} compares with observations.
We conclude with a summary of our findings in Section~\ref{sec:conclusions}.

\section{Methods}
\label{sec:methods}

In this Section we give only a brief overview of our simulation methods and 
subgrid feedback prescriptions, focussing instead on our post-processing. 
Details of the moving mesh implementation in \arepo~may be found 
in \cite{Springel:2010}, while the TreePM approach used to compute gravitational interactions
is described in \cite{Springel:2005}. The framework for supernova and AGN feedback, star formation
and metal enrichment within which our simulations are run is presented fully in V13.
All our analysis scripts are publicly available at \url{https://github.com/sbird/DLA_script}, 
implemented in Python and C++.


\subsection{Parameters of the Simulations}
\label{sec:feedback}

\subsubsection{Supernova Feedback}

Supernova feedback in our simulations, following V13 and \cite{Springel:2003}, is modelled
by the injection of kinetic energy from star-forming regions into the surrounding gas.
This is implemented by stochastically spawning wind particles from gas elements 
with $\rho \geq 0.13\,{\rm  cm}^{-3}$, the star formation threshold density.
Wind particles interact gravitationally, but are hydrodynamically decoupled.
Once they reach a lower density region or a travel time threshold, their energy, momentum, 
mass and metal content are added to the gas cells at their current locations.

\begin{table}
\begin{center}
\begin{tabular}{|l|c|c|c|l|}
\hline
Name & $v^m_w$ (\kms) & AGN & $\kappa_\mathrm{w}$ & Notes \\
\hline 
DEF     &  0       & Yes & $3.7$ & As V13 \\  
WMNOAGN   &  0       & No &  $3.7$ & Warm winds \\ 
HVEL    &  600     & Yes &  $3.7$ & \\ 
HVNOAGN    &  600     & No &  $3.7$ & \\  
2xUV     &  0     & Yes &  $3.7$ & $2\times$ UVB amplitude \\ 
FAST     &  0     & Yes &  $5.5$ & \\ 
NOSN    &  -       & No &  - & No feedback \\ 
\hline
\end{tabular}
\end{center} 
\caption{Simulation parameters varied. $v^m_w$ is the minimum wind velocity
 and $\kappa_\mathrm{w}$ is the coefficient between local velocity dispersion and wind velocity.}
\label{tab:simulations}
\end{table}

Table \ref{tab:simulations} summarises the parameters of our feedback model. DEF is the reference simulation of V13, 
while our other simulations change the parameters listed. As a check on the effects of feedback, we include a simulation 
with metal enrichment but lacking supernova or AGN feedback.
In our other simulations, the wind energy per unit stellar mass is a function of the wind velocity, $v_\mathrm{w}^2$, and mass loading, $\eta_\mathrm{w}$
\begin{align}
  \mathrm{egy}_\mathrm{w}  &=  \frac{1}{2} \eta_\mathrm{w} v_\mathrm{w}^2 \\
			   &= 1.09\; \mathrm{egy}_\mathrm{w}^0 
\end{align}
where $\mathrm{egy}_\mathrm{w}^0 = 1.73 \times 10^{49} \mathrm{erg}\, M_\odot^{-1}$ is the expected available supernova energy per stellar 
mass.\footnote{Note that the values of these parameters are incorrectly listed in V13; see \cite{Vogelsberger:2014e}.}
In order to obtain the large mass loadings necessary to adequately suppress star formation in
low-mass halos \citep{Okamoto:2010, Puchwein:2012}, $v_\mathrm{w}$ is allowed to depend linearly on the local DM 
velocity dispersion, $\sigma^\mathrm{1D}_\mathrm{DM}$. Thus we have 
\begin{align}
  v_\mathrm{w} &= \kappa_\mathrm{w} \sigma^\mathrm{1D}_\mathrm{DM}\,.
  \label{eq:windvel}
\end{align}
\cite{Oppenheimer:2008} showed that this correlates with the virial velocity of the host halo;
smaller halos thus produce winds at a lower velocity and so a larger mass loading. We consider two wind speeds; 
$\kappa_\mathrm{w} = 3.7$, following V13, and, to test the effect of faster winds, $\kappa_\mathrm{w} = 5.5$.
The latter produces a wind velocity mid-way between the default model and the fast winds simulation of V13, 
in which star formation at $z=0$ is almost completely suppressed.

To evaluate the extent to which the increased mass loading in small halos is affecting the DLA population, 
we run a simulation where the winds have a minimum velocity of $600$ \kms. 
This keeps the wind velocity constant for the halos which dominate the DLA cross-section
and increases it when compared to our default model. We also considered a minimum wind velocity of $200$ \kms, and found that it produced an effect 
similar to changing the UVB amplitude, described below.

Following \cite{Marinacci:2013} we considered depositing $40\%$ of the supernova energy in thermal, 
rather than kinetic, energy, producing warmer winds. This had negligible effect on the simulated DLA population.
It is likely that within the high density gas which produce DLAs the extra thermal energy is quickly dissipated by cooling.


In order to match the mass-metallicity relation, V13 allow the wind metal loading to vary 
independently of the wind mass loading, so that the wind metallicity is given by
$Z_\mathrm{w} = \gamma_\mathrm{w} Z_\mathrm{ISM}$. We adopt $\gamma_\mathrm{w} = 0.4$, following V13.
We have checked explicitly by running a simulation with $\gamma_\mathrm{w} = 0.7$ that none of our results 
are sensitive to this parameter.

\subsubsection{AGN Feedback} 

AGN feedback is implemented following \cite{Springel:2005f} and \cite{Sijacki:2007}. 
Massive halos contain accreting black hole particles, which affect their surroundings in three different ways; 
quasar-mode, radio mode and by suppressing cooling. The radio mode is the only one which has a significant 
effect on the surrounding galactic environment. Here thermal energy is periodically dumped into the gas around 
the black hole, allowing sufficient energy to be released in a single burst to overcome the short cooling times of the dense gas.
We will show that the effect of AGN feedback on the DLAs is in most circumstances small, so we will not 
explore the parameter space of the AGN model further in this work.

\subsubsection{Cooling and Enrichment} 

We include radiative cooling implemented using a rate network, including line cooling, free-free emission 
and inverse Compton cooling off the cosmic microwave background \citep{Katz:1996}. 
The UV background (UVB) in most of our simulations follows the estimates of \citet{Faucher:2009},
which was chosen to match constraints from the mean temperature and opacity in the \Lya forest \citep{Faucher:2008c,Faucher:2008d}. 
However, \cite{Becker:2013} find a photo-ionisation rate larger 
by a factor of two, due primarily to an improved measurement of the mean IGM temperature \citep{Becker:2010}, 
rather than a change in the measured mean optical depth. We perform one simulation which doubles the amplitude of the UVB
while leaving its evolution unchanged. This produces a photo-ionisation rate 
in reasonable agreement with the results of \cite{Becker:2013}. We assume a temperature-density parameter 
$\gamma = 1.56$, which matches the Doppler widths of \Lya lines \citep{Rudie:2012, Bolton:2013}, and is roughly the same as that 
produced by the simulations\footnote{$\gamma$ is defined by fitting the IGM equation of state with $\rho \propto T^{\gamma-1}$}.

The gas in our simulations is enriched by mass return from star particles.
The formation rate and yield of AGB stars, type Ia and type II supernovae are computed for each 
particle using a \cite{Chabrier:2003} IMF, as detailed in V13. Metals are distributed into the gas cells 
surrounding the star using a top-hat kernel with a radius chosen to enclose a total mass equal to $256$ times 
the average mass of a gas element. We have checked that distributing the metal within a radius enclosing $16$ times the gas element mass gives 
the same result.

\subsubsection{Initial Conditions}

All simulations use the same initial conditions, originally generated for V13.
The initial redshift is $z=127$, and a linear theory power spectrum used was
obtained from CAMB with cosmological parameters $\Omega_m = 0.2726$,
$\Omega_{\Lambda} = 0.73$, $\Omega_b = 0.0456$, $\sigma_8 = 0.81$, $n_s = 0.963$
and $H_0 =70.4\, \rm km s^{-1} Mpc^{-1}$ ($h=0.704$). The box size is $25 \Mpch$. 
The initial conditions include $512^3$ dark matter particles
with a particle mass of $7.33 \times 10^6 \Msun$, and $512^3$ gas elements
with an initial mass of $1.56 \times 10^6 \Msun$. The gas elements are refined and derefined
throughout the simulation to keep their mass roughly constant.
The dark matter has a fixed comoving gravitational softening length of $1.0 \kpch$. 

\subsection{Neutral Hydrogen Fraction}
\label{sec:nhifrac}

\subsubsection{Self-Shielding}

At $z < 5$, most gas is in ionisation equilibrium with the UVB and thus highly ionised.
However, DLAs are sufficiently dense that they are able to shield themselves from the ionising photons and remain neutral.
Self-shielding becomes important at a relatively sharp hydrogen density threshold 
$n_\mathrm{H} \sim 10^{-2}$ cm$^{-3}$ \citep[e.g.][]{Haehnelt:1998}. Most neutral gas in DLAs has a density
significantly above the threshold, so they are only weakly affected by the details of the transition \citep{Katz:1996a}.
We use the fitting formula provided by \cite{Rahmati:2013a} to compute the self-shielding correction 
to the photo-ionisation rate as a function of the hydrogen density and gas temperature.
Star forming gas is assumed to have a temperature of $10^4$ K.

Some authors \citep[e.g.][]{Fumagalli:2011, Rahmati:2013b} find that local stellar radiation 
may be significant for absorbers with $\NHI \gtrsim 10^{21}$\NHunit. However, \cite{Pontzen:2010} 
find that the effect is negligible. Stellar sources are surrounded by dense photoionised gas with a 
short recombination time and an uncertain, but small, escape fraction for ionising photons.
Calculating this accurately involves physics on parsec scales, far below the resolution of our simulations.
We therefore neglect the effect, noting that it can be viewed as part of the unresolved physics included in our
feedback prescription.

\subsubsection{Molecular Hydrogen}

We include a prescription for the formation of molecular hydrogen, 
following \cite{Altay:2011}, based on the relationship between the 
surface density of molecular hydrogen and the mid-plane hydrostatic 
pressure found by \cite{Blitz:2006}. Here the pressure from the effective 
equation of state in star-forming gas is used as a proxy for the hydrostatic pressure. 
Non star-forming gas does not form molecular hydrogen.
The molecular fraction in star-forming gas is given by:
\begin{equation}
 f_\mathrm{H2} = \frac{\Sigma_\mathrm{H2}}{\Sigma_\mathrm{H}} = \frac{1}{1+   (35 ( 0.1 / n_\mathrm{H} )^{5/3})^{0.92}}\,,
\label{eq:molecular}
\end{equation}
where $n_\mathrm{H}$ is the hydrogen density in cm$^{-3}$.

\cite{Blitz:2006} measured molecular fractions for systems with metallicities as low as $0.1 \Zsun$, and found no 
significant dependence on metallicity, although this is expected theoretically
\citep[e.g.][]{Schaye:2001}. \cite{Erkal:2012} suggested that this relation therefore may not hold in 
the metal-poor dwarf galaxies that make up DLAs in their simulations.
We checked the metallicity histogram of the star forming cells, weighted by H$_2$ mass,
in our simulation and found that $85\%$ of the H$_2$ was in particles with metallicity $ Z > 0.1 \Zsun$.
This suggests that any suppression of H$_2$ formation at low metallicity would affect the total amount of H2 formed at the $10\%$ level at most, 
giving negligible effect on the column density distribution function (CDDF). Furthermore, \cite{Fumagalli:2010} 
found that on the scales relevant to our simulations the \cite{Blitz:2006} relation is 
valid even for very metal-poor gas.

\subsection{Projection and DLA Identification}
\label{sec:colden}

The rarity of high density columns (due to our small box size, $< 1$\% of our spectra will contain a DLA) 
necessitates a large number of sightlines to collect a representative sample of DLAs.
We project all particles in the volume onto a regular grid covering the entire box, 
treating each grid cell as a sightline. Each grid cell has a linear size of $1.5 \kpch$, approximately the gravitational softening length, 
and thus the projected box contains $16384^2$ cells.

Gas elements are projected onto the grid using an SPH kernel. The SPH smoothing length is chosen so that the kernel support 
equals the volume of the moving mesh gas cell. Should multiple DLAs be present in the same cell, but at different depths along the projection axis, 
they may be unphysically blended together.
We checked that for a $25 \Mpch$ box this did not have a significant affect on the statistics of DLAs.
However, it did become difficult to distinguish the boundaries of systems visually, as projection effects caused some large systems 
to appear close together. Therefore, we computed the projected neutral hydrogen column density in ten projected slices of $2.5 \Mpch$ width. 
This treats each cell as a single quasar sightline containing ten pixels evenly spaced across the box.
We verified that our results were converged with respect to both the grid cell size and the slice width at the $5\%$ level.
The column density, for a projection along the $x$ axis, is thus
\begin{equation}
 N_\mathrm{HI} =  \frac{(1+z)^2}{m_\mathrm{P}} \int {\rm d}x\, \rho_\mathrm{HI}(x)\,,
\label{eq:column_density}
\end{equation}
where $m_\mathrm{P}$ is the proton mass and $\rho_\mathrm{HI} (x)$ 
is the neutral hydrogen density. The factor of $(1+z)^2$ enters because $\rho_\mathrm{HI}$ is 
in comoving units and $N_\mathrm{HI}$ is in physical units.
A DLA is defined to be an absorber with $N_\mathrm{HI} > 10^{20.3}$ \NHunit.

\subsection{DLA Host Halos}
\label{sec:halos}

\begin{figure*}
\includegraphics[width=0.45\textwidth]{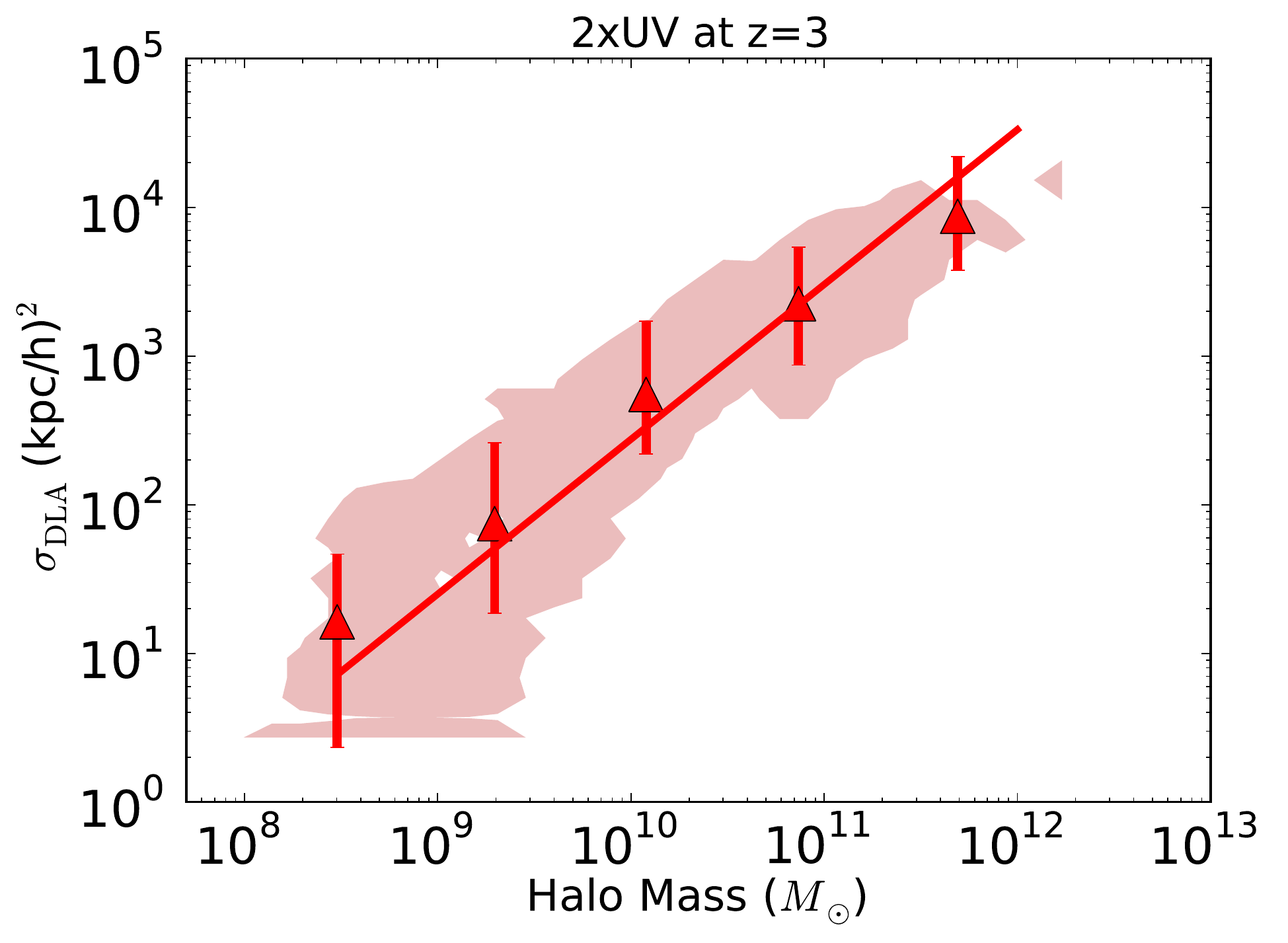}
\includegraphics[width=0.45\textwidth]{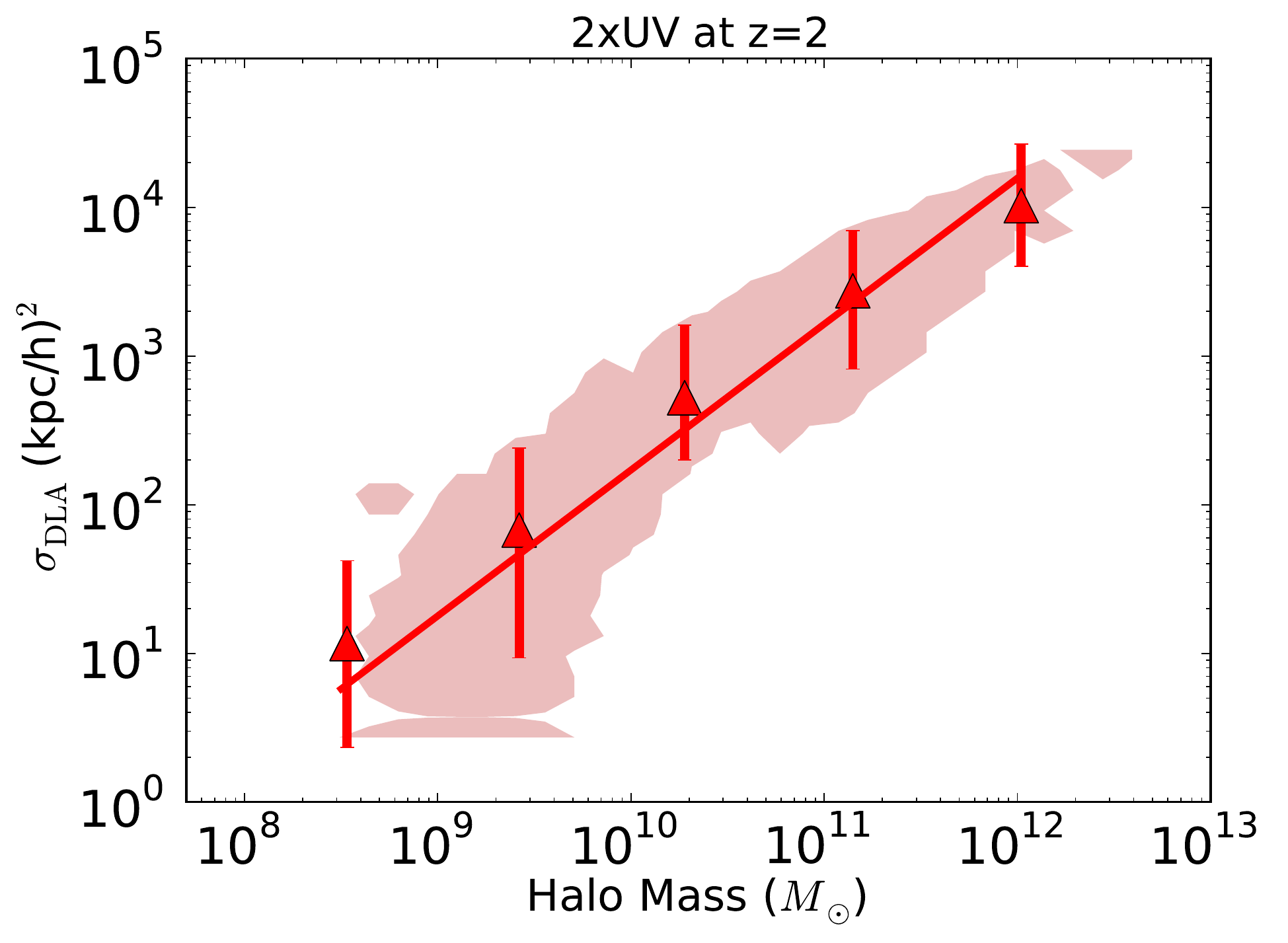}
\caption{The cross-sectional area of DLAs around an individual halo, $\sigmaDLA$, as a function of halo mass for the 2xUV simulation. Shaded areas show regions of the plane 
occupied by at least one halo. Points indicate the median value of $\sigmaDLA$ in comoving $(\kpch)^2$ in each mass bin, and error bars give the upper and lower quartiles.
The solid line is a power law fit. (Left) At $z=3$. (Right) At $z=2$.}
\label{fig:dla_halos}
\end{figure*}

\begin{figure*}
\includegraphics[width=0.45\textwidth]{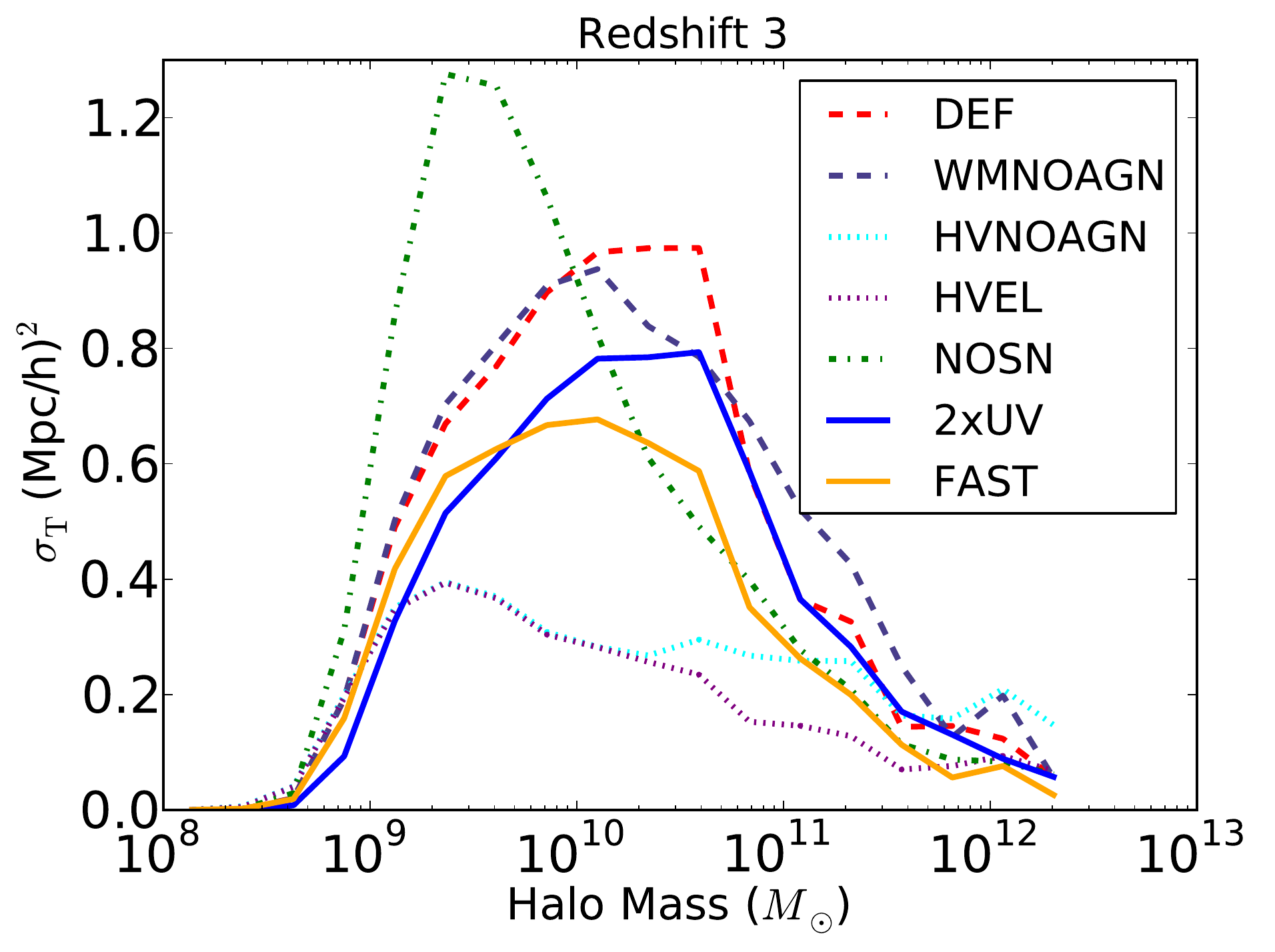}
\includegraphics[width=0.45\textwidth]{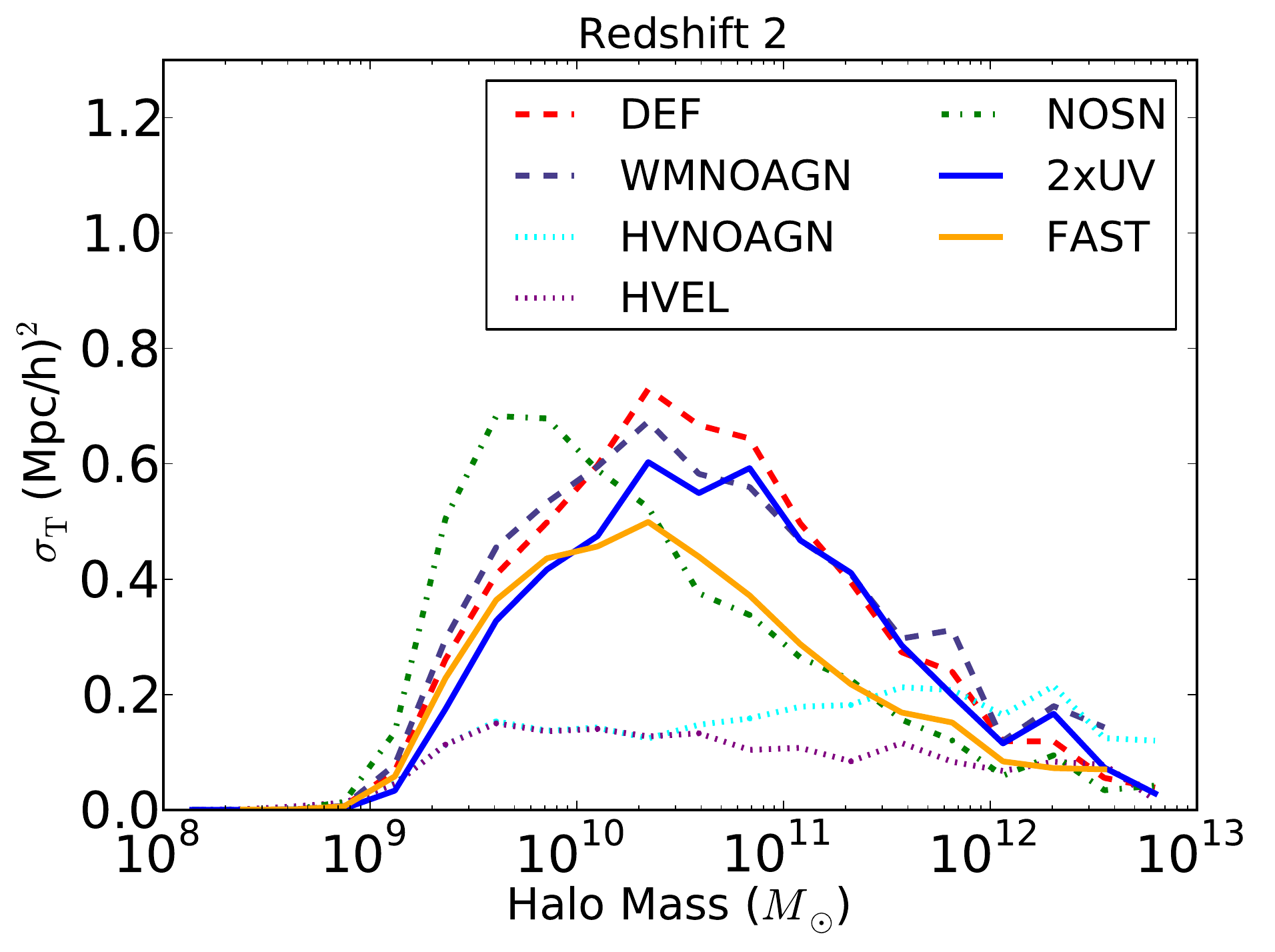}
\caption{$\sigma_\mathrm{T}$, the total cross-section of DLAs, in comoving units, hosted by all halos in each bin as a function of halo mass.
Shown are $20$ halo mass bins evenly spaced in log$(M)$, for a selection of our simulations. (Left) At $z=3$. (Right) At $z=2$.
}
\label{fig:dla_halohist}
\end{figure*}

DLAs are typically associated located within a host halo. 
To identify this halo, we must first find the position of each DLA in the projection direction.
This we define as the HI weighted average depth of the particles within each pixel.
For reasons of speed and numerical stability, we use only particles with a neutral fraction greater than $10^{-3}$.
If the DLA lies within the virial radius of a halo, it is assigned to that halo.
If a DLA instead lies within the half-mass radius of a subhalo, which is the case for $5-10$\% of DLAs, we assign 
it to that subhalo's parent. The $\sim 5$\% of DLAs which are not within the virial radius of any halo or subhalo 
are classed as field DLAs.

\subsection{DLA Bias}
\label{sec:dlabias}

The linear DLA bias, $b_\mathrm{DLA}$, is defined as
\begin{equation}
P_\mathrm{DLA} (k,z) = b^2_\mathrm{DLA}(k,z) P_\mathrm{M} (k,z)\,,
\label{eq:dlabias}
\end{equation}
where $P_\mathrm{DLA}$ is the DLA power spectrum and $P_\mathrm{M}$ the matter power spectrum.
Observationally, $b_\mathrm{DLA}$ is measured by cross-correlating DLAs with other tracers, such as 
the \Lya forest \citep{FontRibera:2012}, or Lyman-break galaxies \citep{Cooke:2006}.
Because strong absorbers are rare, cross-correlating with more common objects increases the statistical power of the measurement. 
In our simulated results, however, we can produce arbitrarily dense quasar sightlines, so we compute $b_\mathrm{DLA}$ 
directly from Eq. (\ref{eq:dlabias}) (the autocorrelation). $P_\mathrm{M}$ is the total matter power spectrum computed from the 
simulation box, rather than the large-scale linear power.

Both $P_\mathrm{DLA}$ and $P_\mathrm{M}$ are computed using a Fourier mesh with $4096^3$ cells.
Gas elements are interpolated onto the mesh using an SPH kernel, while dark matter and star particles are interpolated with 
a cloud-in-cell kernel. The DLA power spectrum is calculated from the interpolated gas density grid by computing
\begin{align}
\delta_\mathrm{DLA} &= \mathrm{I}( \mathrm{DLA} ) / \bar{n}_\mathrm{DLA} - 1\,,
\label{eq:deltadla}
\end{align}
where $\mathrm{I}( \mathrm{DLA} )$ is unity when a cell contains a DLA and zero otherwise.
Here $\bar{n}_\mathrm{DLA}$ is the average density of DLAs per cell and we define $P_\mathrm{DLA}$ as the power spectrum of $\delta_\mathrm{DLA}$.
This procedure will introduce some smoothing on the scale of the Fourier mesh, causing some DLAs to be misclassified. We checked the size of this effect by comparing with a $2048^3$ grid, 
and found that the DLA bias was converged to $\sim 3$\% percent.
We compute the bias at $z=2.3$ by linearly interpolating log $b_\mathrm{DLA}(k)$ as a function of scale factor, $a$, between $z=2.5$ and $z=2$.

\section{DLA Properties}
\label{sec:properties}

\subsection{Host Halos of DLAs} 
\label{sec:halosize}


\begin{table}
\begin{center}
\begin{tabular}{|l|c|c|}
\hline
Name & $\sigma_{10}$ ($\kpch^2$) & $\beta$ \\
\hline 
DEF	&   2.52   & 1.01   \\
WMNOAGN   &  2.58    & 1.06   \\ 
HVEL    &  2.27    & 0.81   \\ 
HVNOAGN &  2.36    & 0.91   \\  
2xUV     &  2.44    & 1.04 \\  
FAST     &  2.40    & 0.93     \\ 
NOSN    &  2.53     & 0.81  \\ 
\hline
\end{tabular}
\end{center} 
\caption{Table of best-fit parameters to Eq. (\ref{eq:powerlaw}) at $z=3$. 
}
\label{tab:params}
\end{table}
Figure \ref{fig:dla_halos} shows the cross-sectional area of DLAs in halos, $\sigmaDLA$, for the 2xUV simulation at
$z=3$ and $z=2$. There is a cut-off at small halo masses where the DLA cross-section becomes negligible. The position of this cut-off is 
set by the amplitude of the UVB, which prevents gas in halos with a low virial velocity from collapsing 
enough to become self-shielded \citep{Okamoto:2008}. 

We have fit $\sigmaDLA$ with a power law
\begin{equation}
 \sigmaDLA = \sigma_{10} \left(\frac{M}{10^{10}}\right)^{\beta}\,.
\label{eq:powerlaw}
\end{equation}
Table \ref{tab:params} lists the parameters of this fit for all our simulations at $z=3$.
The fit was carried out to the median binned $\sigmaDLA$ in each halo mass bin.
We used only bins with $M > 10^{8.5} \Msun$, which roughly corresponds to the cut-off in the DLA cross-section - halo mass relation.
At higher halo masses the relation flattens \citep{Barnes:2009}, and so the power law we propose should not be 
extrapolated to higher halo masses than in our simulation. The parameters of the power law relation for simulation 
2xUV at $z=4$ are $2.86$ and $1.02$, and at $z=2$ are $2.19$ and $1.00$.
At lower redshift more concentrated halos and an increased UVB amplitude produce a smaller DLA cross-section per host halo.
However, the slope of the relation, which is remarkably linear, exhibits minimal evolution with redshift, although it extends 
towards higher masses as halos grow.

Figure \ref{fig:dla_halohist} shows the total cross-section of DLAs in each mass bin, summed over all halos.
While this is computed directly from the simulation, it can be thought of as a convolution of the halo mass function with Eq.~(\ref{eq:powerlaw}).
More massive halos have a greater DLA cross-section per halo, but are less common, and so have a lower total cross-section. 

For the NOSN simulation, the lack of outflows means that the slope of the DLA cross-section is shallow and the host 
halo mass distribution peaks around $10^{9} \Msun$, below which the UVB prevents gas in halos from condensing. 
However, supernova feedback prevents the runaway collapse of small halos into DLAs, significantly reducing their cross-section and increasing
the slope of the halo mass-$\sigmaDLA$ relation. The HVEL and HVNOAGN simulations have the flattest DLA distribution; here the high wind velocity suppresses DLAs in host halos 
with $M < 10^{12} \Msun$. For the other models, which we shall see agree best with observations, the peak halo mass is at $5 \times 10^{10} \Msun$. 
The higher velocity winds in the FAST simulation suppress the DLA abundance roughly uniformly across the whole 
range of host halo masses. AGN feedback preferentially suppresses DLAs in high mass halos; WMNOAGN (HVNOAGN) differs from DEF (HVEL) only for halos with $M > 10^{10.2} \Msun$. 
We verified that it reduces the DLA cross-section in these halos, rather than reducing the number of halos which host DLAs. 

\begin{figure*}
\includegraphics[width=0.45\textwidth]{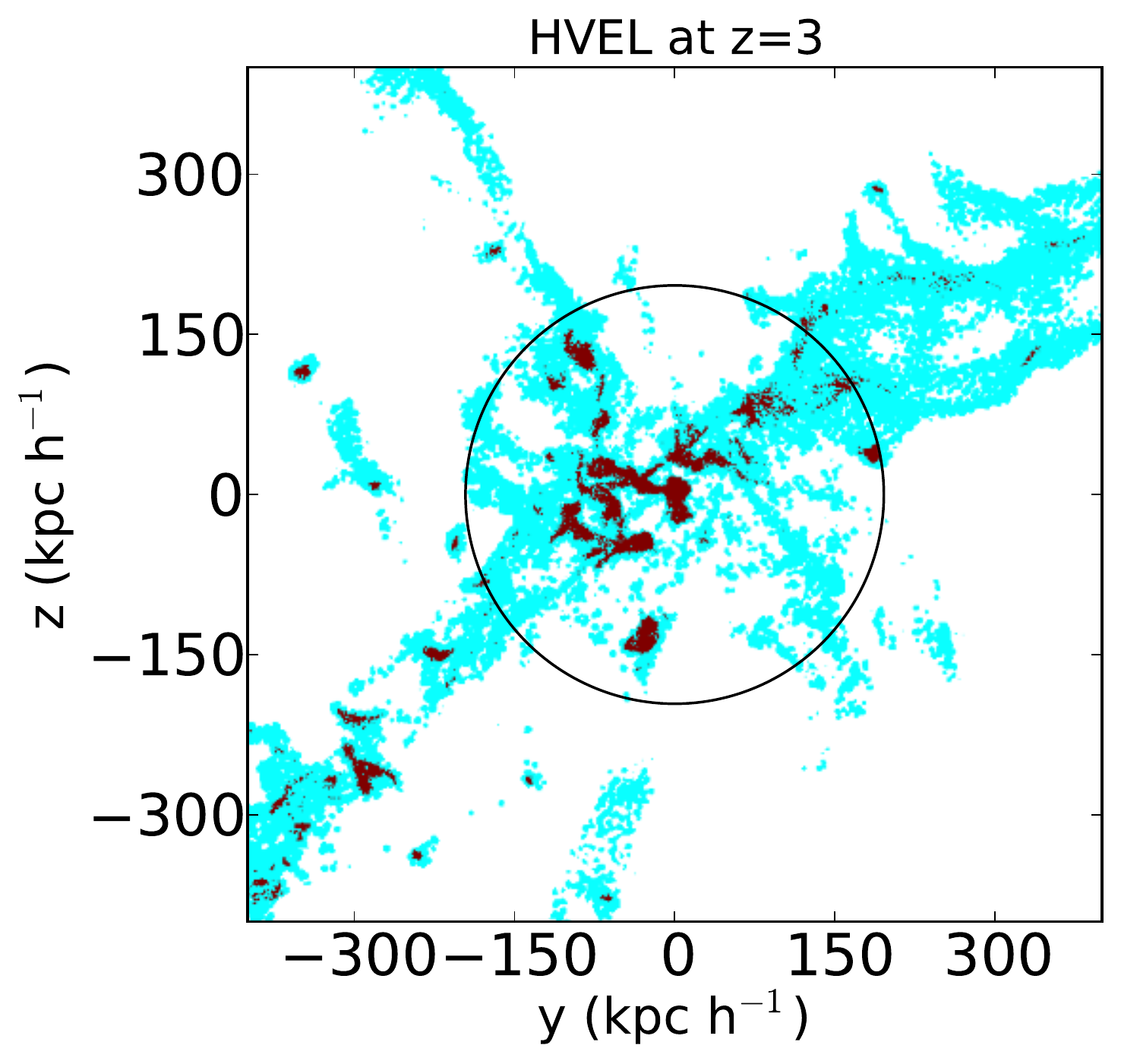}
\includegraphics[width=0.45\textwidth]{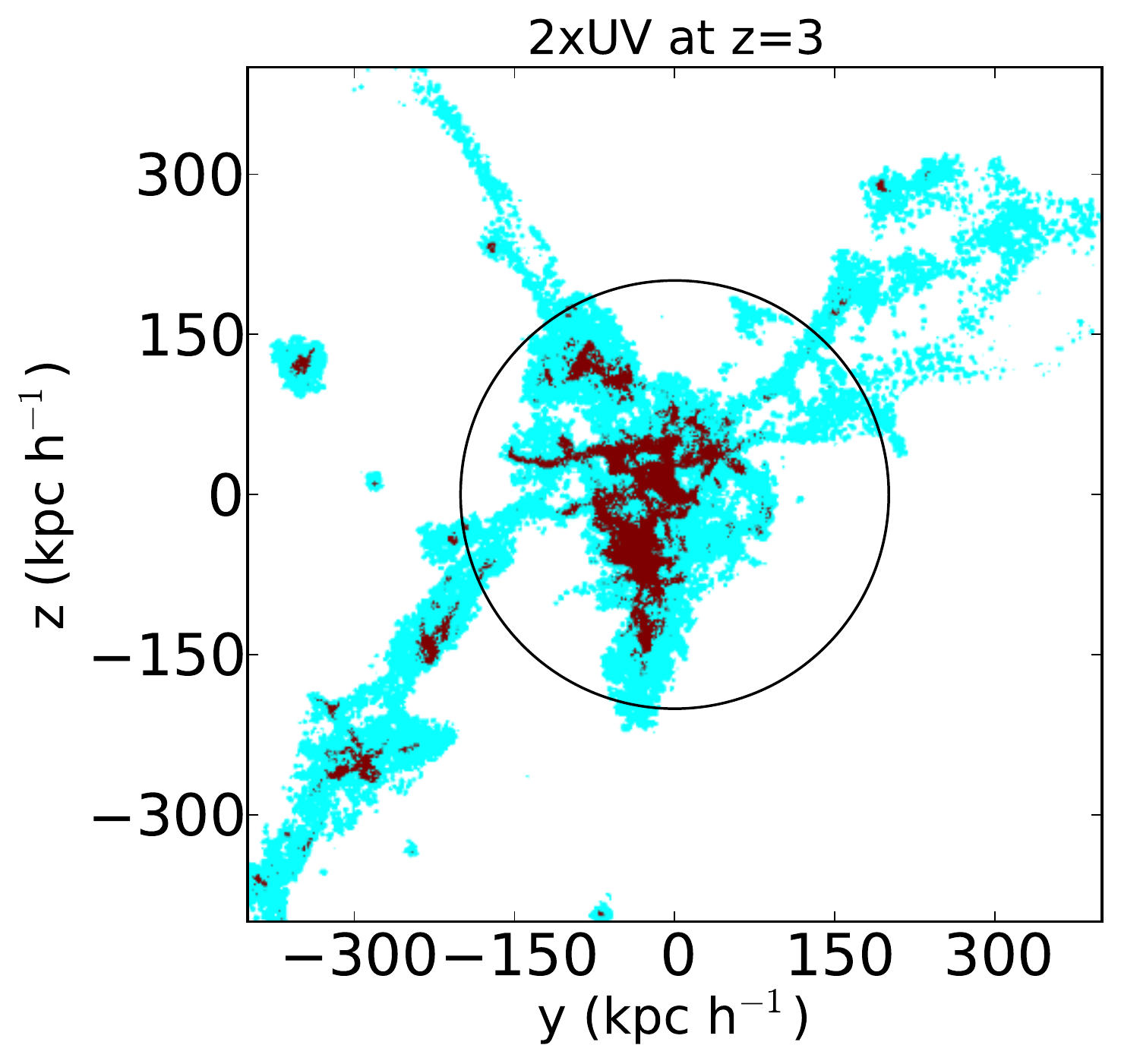}
\caption{Distribution of DLAs and LLSs around a typical DLA hosting halo, with a central halo mass of $\sim 7\times 10^{11}\Msun$.
DLAs are shown in dark red, LLSs in light blue. Distances are comoving. 
The black circle shows the virial radius of the central halo. (Left) From the HVEL simulation. (Right) From the 2xUV simulation. }
\label{fig:pretty_dla_halos}
\end{figure*}

Figure \ref{fig:pretty_dla_halos} shows visually the distribution of DLAs and LLSs around a typical halo in two of our simulations, HVEL and 2xUV. Halos
have a central density of neutral hydrogen connected with their central galaxy, surrounded by smaller concentrations associated with subhalos, 
merging objects or the denser parts of filaments. In 2xUV the slower outflows produce a significantly more extended central component than in HVEL.

\subsection{Convergence}
\label{sec:converge}



A $10 \Mpch$ box was used to check convergence with resolution.
This box was too small to include a statistically significant sample of halos with $M \gtrsim 10^{11}\Msun$, which, 
as shown in Figure \ref{fig:dla_halohist}, contribute significantly to the total DLA density.
Thus it did not completely reproduce the DLA population. However, we checked that the DLA cross-section of halos 
as a function of halo mass was the same in both simulations, and that DLA halos below the mass scale resolved in 
our $25 \Mpch$ simulation made up less than $1\%$ of the total DLA cross-section.
We also checked that the total DLA cross-section in halos with $M = 10^9-10^{10} \Msun$ did not change.
Note, however, that a $25 \Mpch$ box with $256^3$ particles is not converged.

To check convergence with box size, we computed the column density function for 
the recently completed Illustris simulation \citep{Vogelsberger:2014, Vogelsberger:2014b, Genel:2014}; a $75 \Mpch$ box at similar resolution to the simulations used here.
Convergence is remarkable; the CDDF was unchanged except for a small region at columns of $4\times 10^{20}$, where it was larger by about $20\%$. 
We attribute this mostly to the slightly reduced grid resolution we were forced by memory limitations to use for the larger box.
Evolution due to improved statistics in the most massive halos would be concentrated towards higher columns.

\section{Comparison With Observations}
\label{sec:results}

\subsection{Column Density Distribution Function}
\label{sec:cddf}

\begin{figure*}
\includegraphics[width=0.48\textwidth]{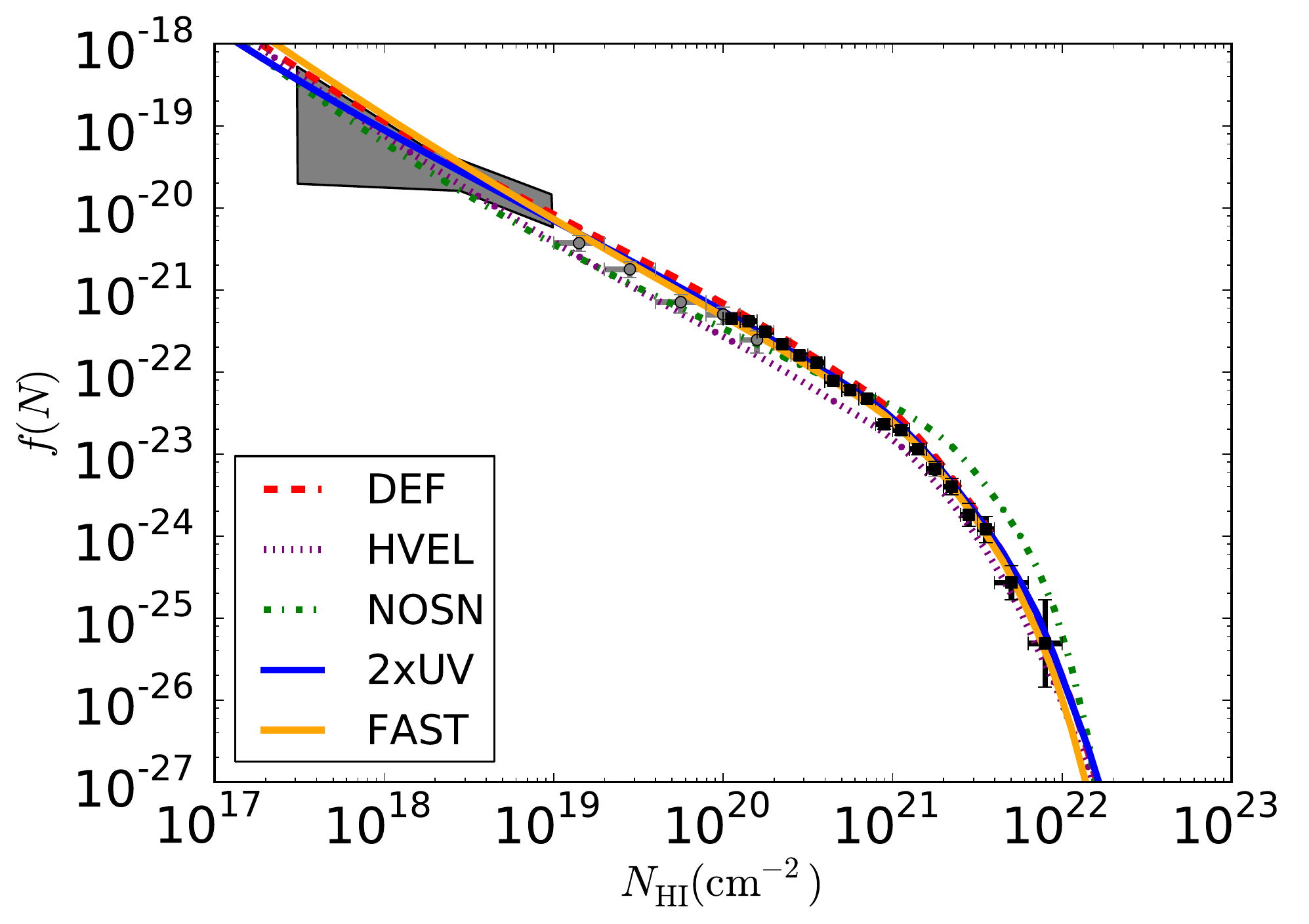}
\includegraphics[width=0.47\textwidth]{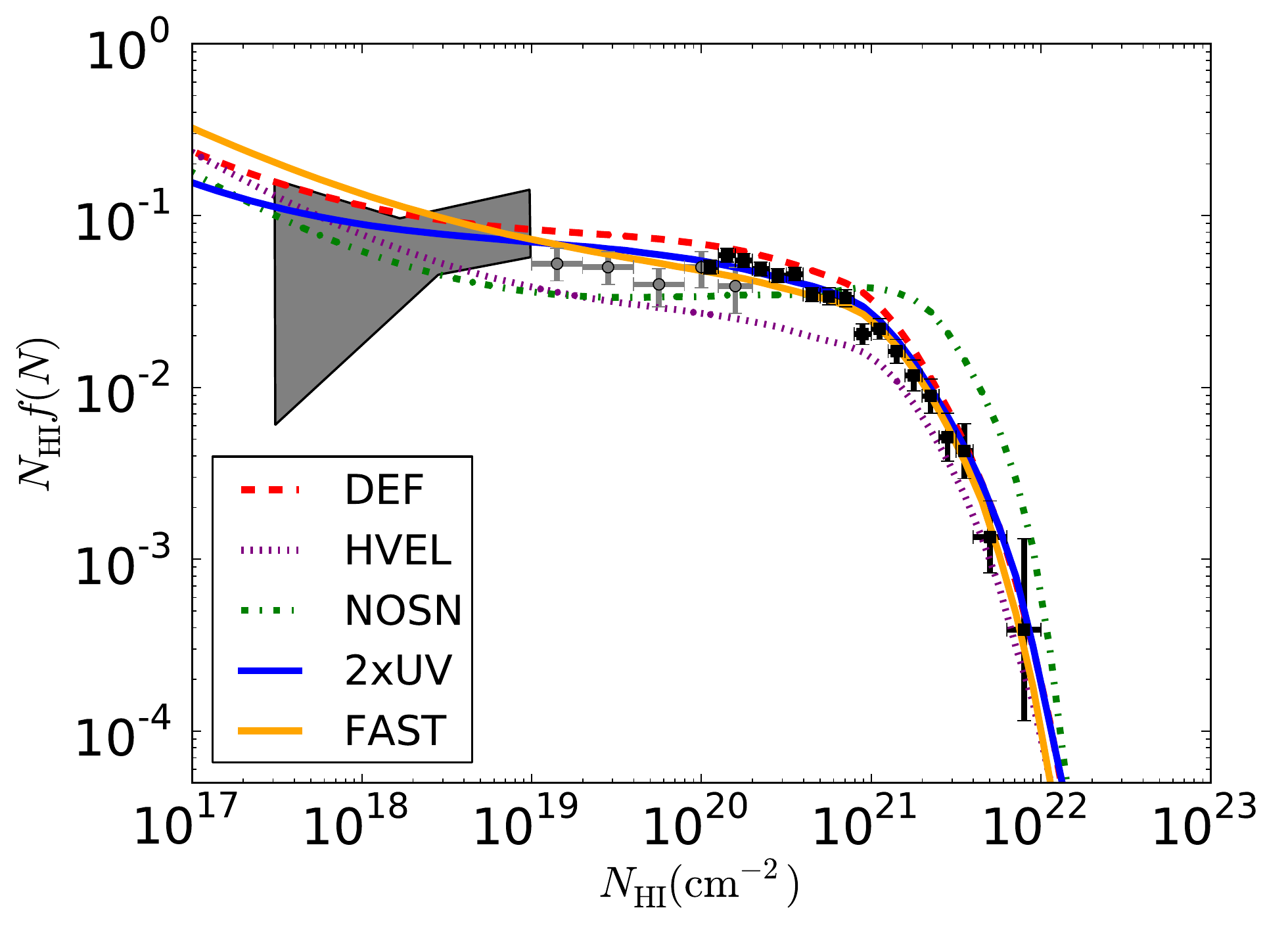}
\caption{The column density distribution function for our simulations at $z=3$, compared to
observational data from \protect\cite{Noterdaeme:2009} at $z=2.9$ (black squares), \protect\cite{Zafar:2013} (grey circles) and \protect\cite{Prochaska:2010} (grey box).
(Left) $f(\NHI)$ as a function of column density. (Right) The first moment of the column density, $\NHI f(\NHI)$.
$\NHI$ is in physical cm$^{-2}$.
}
\label{fig:cddf_feed}
\end{figure*}

The column density distribution function (CDDF), $f(\NHI)$,
is defined such that $f(\NHI)\, {\rm d}\NHI\, {\rm d}X$ is the number of absorbers per
sightline with column density in the interval $[\NHI, \NHI + {\rm d}\NHI]$.
We identify sightlines with grid cells and thus count absorbers by computing 
a histogram of the column densities in each cell. The CDDF is defined by 
\begin{align}
 f(N) &= \frac{F(N)}{\Delta N}{\Delta X(z)}\,,
\end{align}
where $F(N)$ is the fraction of the total number of grid cells in a given column density bin, 
and $\Delta X(z)$ is the absorption distance per sightline. The absorption distance is defined to 
account for evolution in line number density with the Hubble flow.
A quantity $Q$ with constant physical cross-section and comoving number density evolving passively with redshift 
will have constant $\mathrm{d}Q/\mathrm{d}X$. The (dimensionless) absorption distance is thus given by
\begin{equation}
 X(z) = \int_0^z (1+z')^2  \frac{H_0}{H(z')} {\rm d}z'\,,
\end{equation}
and for a box of co-moving length $\Delta L$ we have $\Delta X = (H_0 /c) (1+z)^2 \Delta L$ \citep{Bahcall:1969}.

Figure \ref{fig:cddf_feed} shows the CDDF and its first moment, $\NHI f(\NHI)$, which emphasises the differences between simulations,
at $z=3$. Several of our simulations are in reasonable agreement with the data. 
The DEF simulation over-produces weak DLAs and LLSs by about $1 \sigma$. 
This is corrected in the 2xUV simulation, which uses photo-ionisation rates close to those from \cite{Becker:2013}. 
This reduction of the LLS abundance in the 2xUV simulation is due to two physical effects. First, 
the increased UVB amplitude ionises the gas further. Second, there is increased thermal pressure 
support in small halos, reducing the amount of gas that is able to remain cool. We checked 
that both effects contribute by changing the UVB amplitude when post-processing the DEF simulation.
At low column densities, the gas is beginning to come into photo-ionisation equilibrium 
with the UVB, and is only weakly affected by self-shielding, hence the difference between 
2xUV and DEF tends to a constant.
The FAST simulation is similar to 2xUV at columns $\NHI > 10^{18}$ \NHunit, showing a degeneracy between the UVB amplitude 
and the wind velocity. However, at $\NHI < 10^{18}$ \NHunit, the FAST CDDF increases until it is larger than DEF, as the 
faster winds move gas into less dense regions.

A minimum wind velocity decreases the abundance of DLAs, particularly weak DLAs and LLSs.
Two effects, both tending to reduce the HI cross-section, contribute. First, an increased minimum wind velocity 
decreases the mass loading, allowing the gas in equivalent mass halos to become more concentrated, and thus decreasing the spread of neutral gas.
Second, when the wind velocity is much greater than the circular velocity of the halo, $v_\mathrm{w} \gg v_\mathrm{circ}$, out-flowing gas 
can escape the halo entirely and reach lower density regions where it is ionised by the UVB. The minimum wind velocity of 
$600$ \kms in simulation HVEL is large enough to suppress the DLA abundance over the halo mass range which dominates the DLA cross-section. 
It thus behaves similarly to the REF OWLS model \citep{Altay:2013, Rahmati:2013a}, and yields a CDDF with a shape in good agreement with 
the data, but a normalisation a factor of two too low.

\begin{figure}
\includegraphics[width=0.45\textwidth]{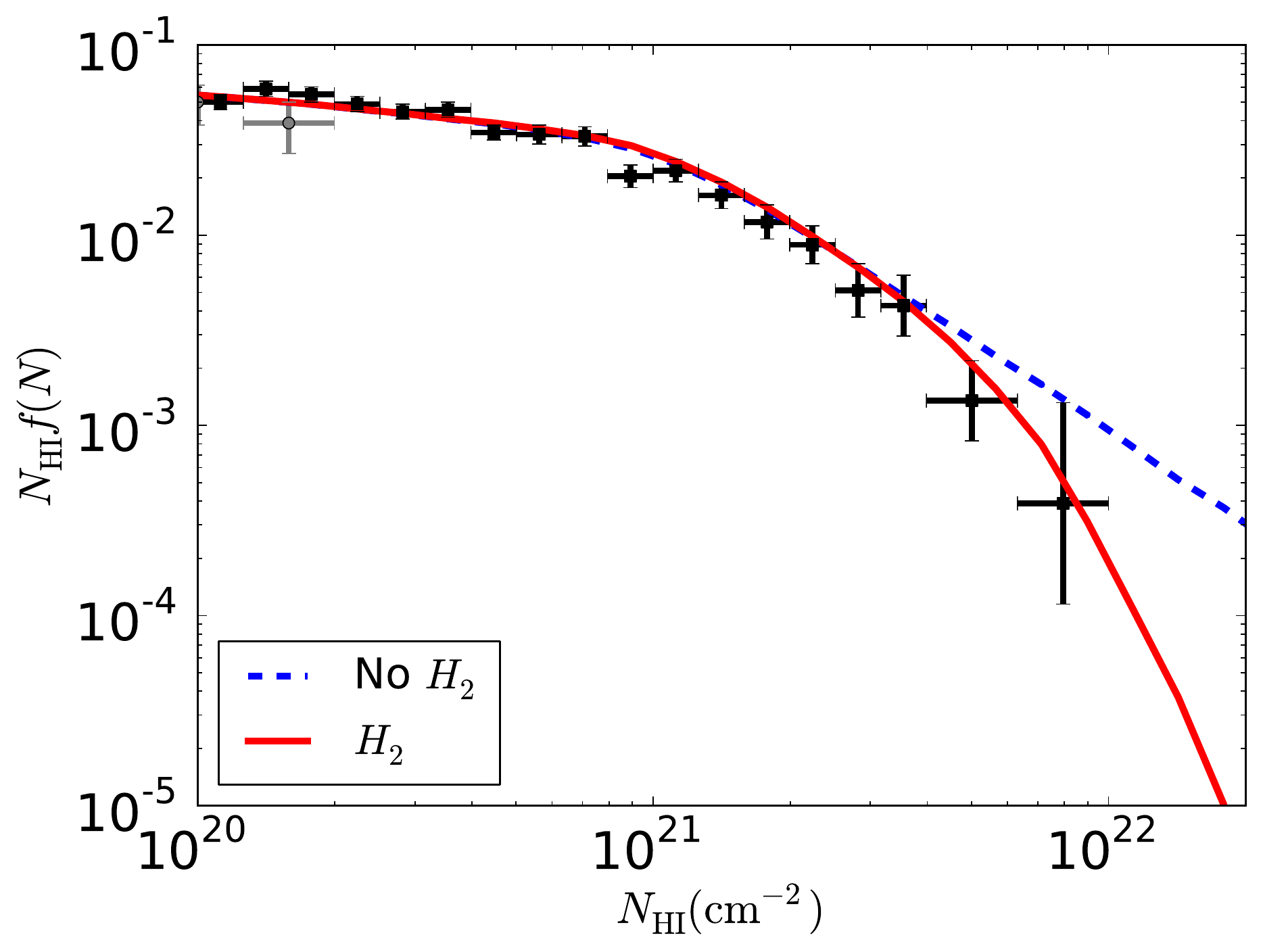}
\caption{The first moment of the CDDF for the 2xUV simulation with and without $H_2$ formation at $z=3$,  
compared to observational data. $\NHI$ is in physical cm$^{-2}$.}
\label{fig:cddf_H2}
\end{figure}

Simulation NOSN, which does not have significant feedback, produces too few LLSs, but too many absorbers with column densities $\NHI > 10^{21}$ \NHunit.
Without feedback, halos undergo runaway gravitational collapse and form dense objects hosting both stars and DLAs. 
Figure \ref{fig:cddf_H2} shows the 
consequences of including H$_2$ formation for the 2xUV simulation. The effect is negligible for 
$\NHI < 10^{21.5}$ \NHunit, but at higher column densities decreases the CDDF by roughly $1 \sigma$. 
Note that even without molecular hydrogen our simulations 
are in reasonable agreement with the data, so any inaccuracies in our prescription for $H_2$ formation 
will not strongly affect our conclusions. 

\begin{figure*}
\includegraphics[width=0.45\textwidth]{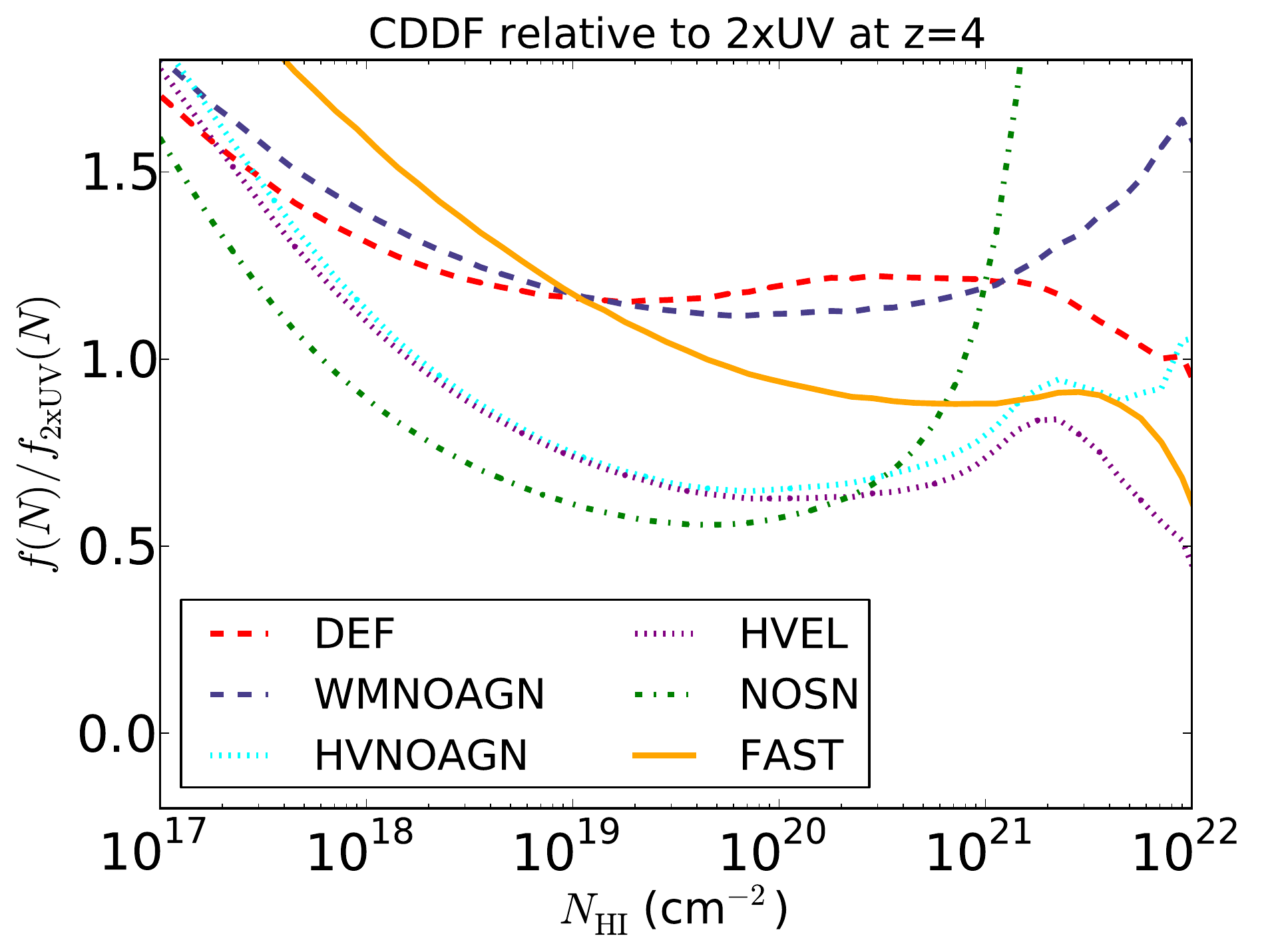}
\includegraphics[width=0.45\textwidth]{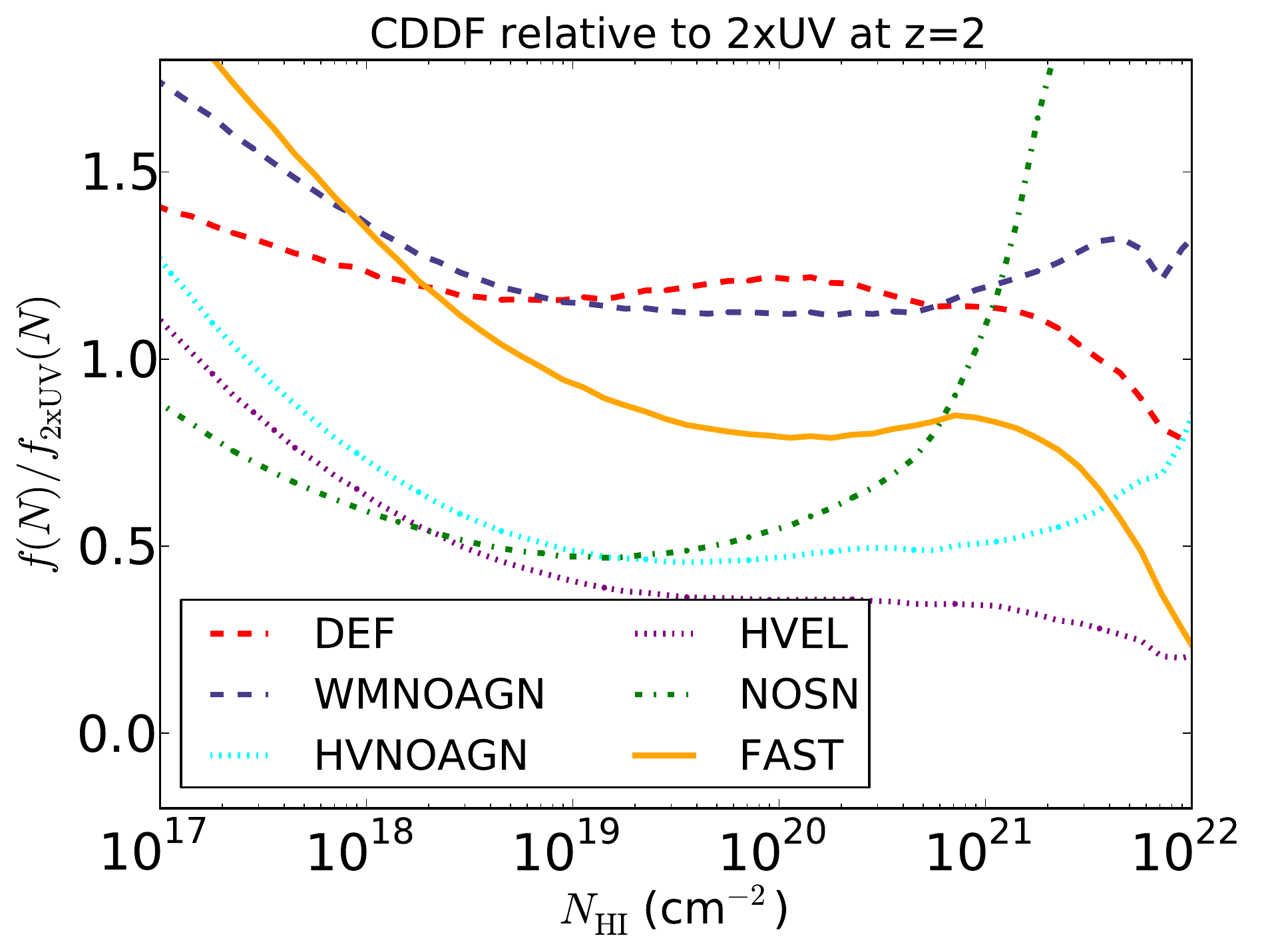} 
\caption{The ratio of the CDDF in our simulations to the CDDF of simulation 2xUV. (Left) At $z=4$. (Right) At $z=2$. }
\label{fig:cddf_feed_rel}
\end{figure*}

Figure \ref{fig:cddf_feed_rel} shows the differences between the simulations at $z=4$ and $z=2$ in more detail. 
The effect of AGN feedback is small, but most important at $z=2$, where there are more large halos.
AGN feedback is more effective in HVEL than in DEF, as a larger fraction of the DLAs in HVEL reside in halos massive enough to host AGN.
Furthermore, the stronger supernova feedback in HVEL produces lower density gas on the outskirts of halos with a longer cooling time and 
thus a larger response to AGN heating. Note that the property of HVEL which increases the effect of AGN also 
causes it to disagree with observations.

It is interesting to compare Figure \ref{fig:cddf_feed_rel} to Figure 2 of \cite{Altay:2013}, 
who considered the effect of a range of feedback and star formation parameters on DLAs.
The variation between their models is bounded by $0.2$ dex in the LLS range, 
with the exception of one model where the Universe is not re-ionised (and thus, freed 
of the effect of photo-ionisation, much smaller halos can form DLAs) and one model with a 
different cosmology. This bound is saturated by their WVCIRC and REF models, whose wind mass loading in small halos 
is different by a factor of $\sim 3$. By contrast, the wind mass loading in small halos in our DEF simulation differs from that in the HVEL simulation by 
$\sim 16$, producing a variation in the LLS abundance nearer $0.4$ dex. This emphasises the importance of considering 
a wide feedback parameter space.

\begin{figure*}
\includegraphics[width=0.45\textwidth]{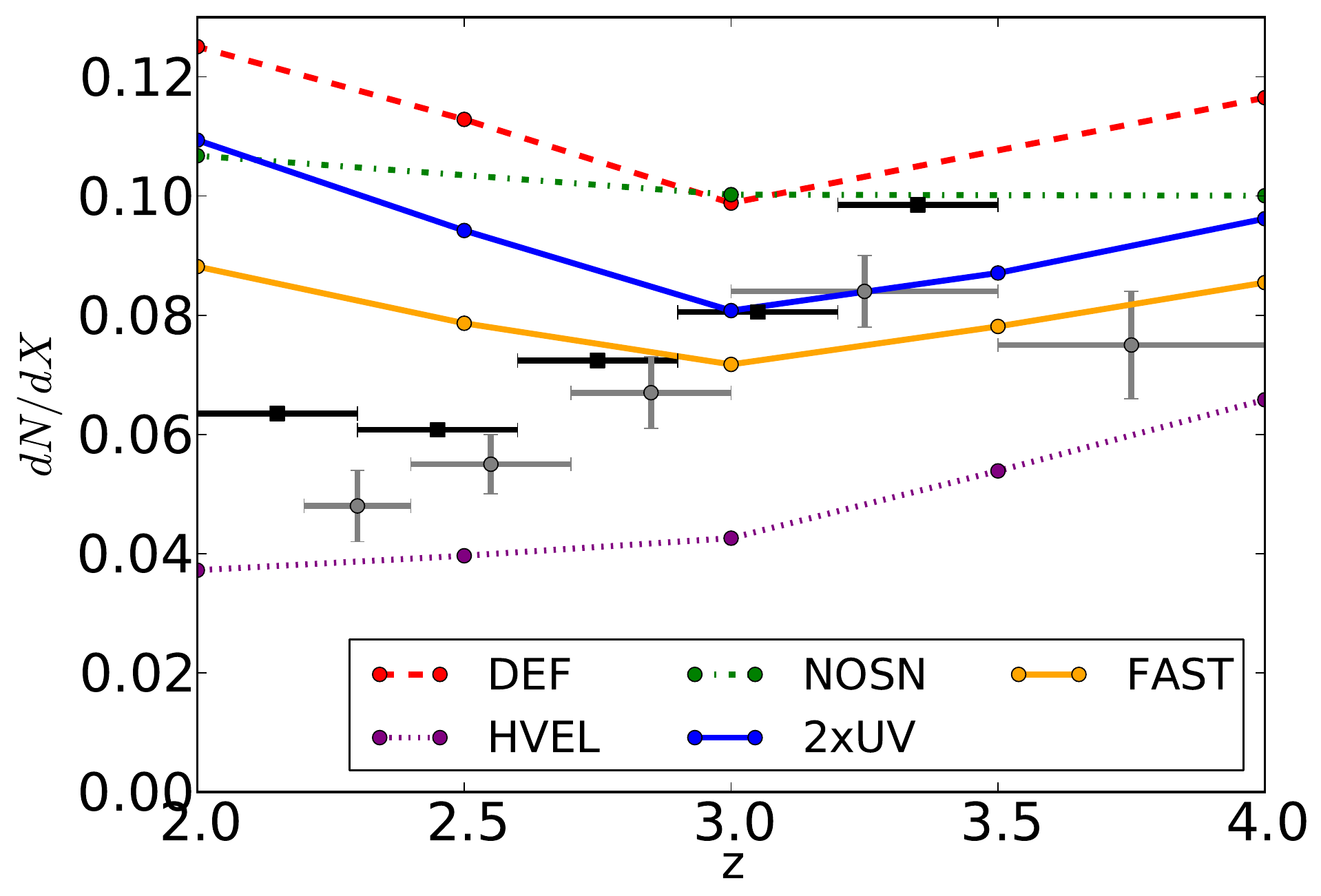}
\includegraphics[width=0.45\textwidth]{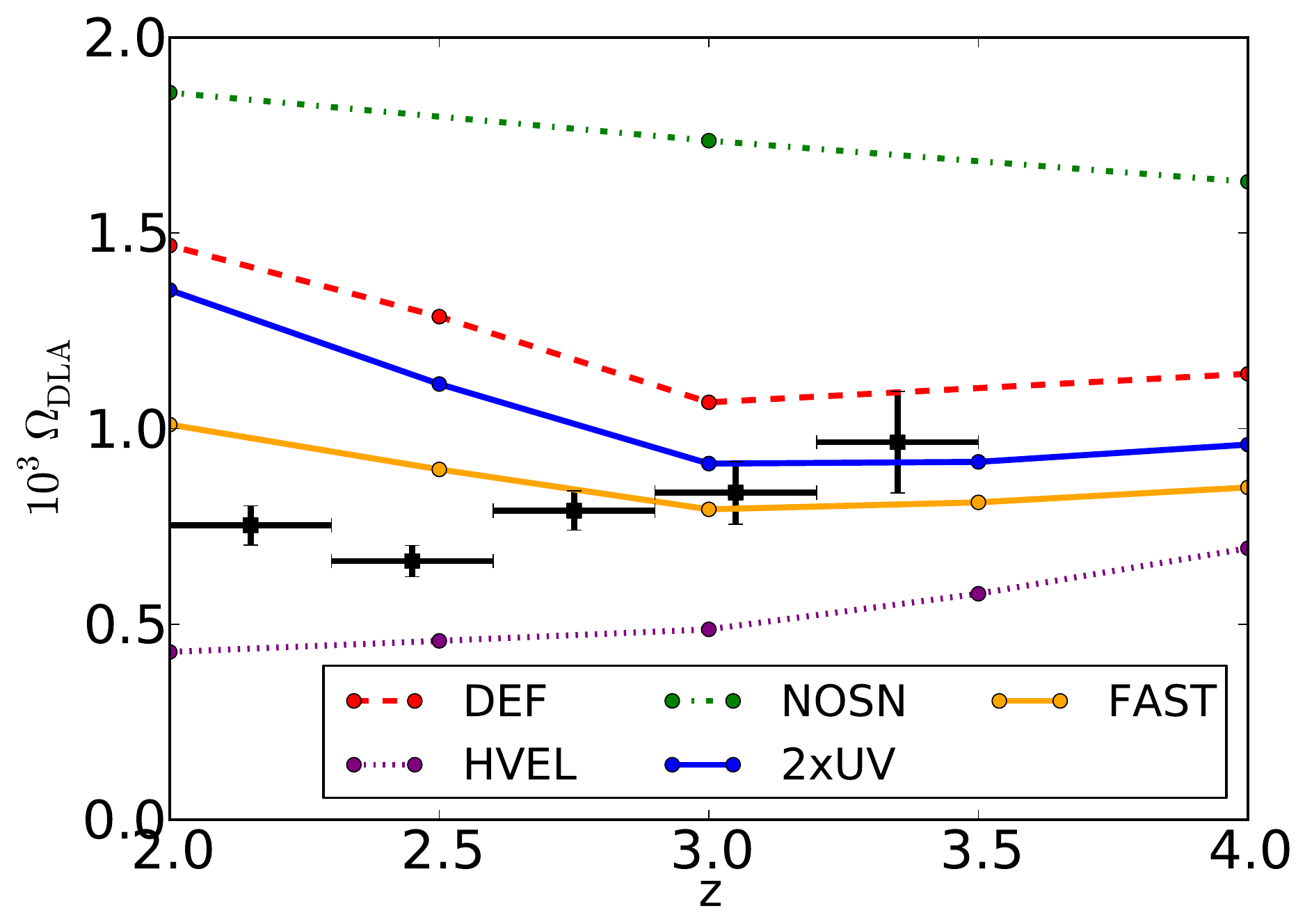}
\caption{(Left) Redshift evolution of the total incident rate of DLAs, $\frac{\mathrm{d}N}{\mathrm{d}X}$. (Right) Total neutral hydrogen density in DLAs, $\Omega_\mathrm{DLA}$. 
Black squares show the observational data of \protect\cite{Noterdaeme:2012}. For $\Omega_\mathrm{DLA}$, errors shown are statistical. 
For $\frac{\mathrm{d}N}{\mathrm{d}X}$, systematic errors in the DLA detection pipeline dominate and so no error is quoted in \protect\cite{Noterdaeme:2012}. 
Thus we show the independent estimate from \protect\cite{Prochaska:2009} (grey circles); the difference between them is a 
reasonable estimate of systematic uncertainty. $\Omega_\mathrm{DLA}$ is dominated by stronger absorbers, and so is less affected.
}
\label{fig:line_dens}
\end{figure*}

%

\subsection{Redshift Evolution of Neutral Hydrogen}
\label{sec:redshift}

Figure \ref{fig:line_dens} shows the redshift evolution of the incident rate of DLA systems, $\frac{\mathrm{d}N}{\mathrm{d}X}$, and the total HI density in DLAs, $\Omega_\mathrm{DLA}$. 
$\frac{\mathrm{d}N}{\mathrm{d}X}$ is the integral of the CDDF
\begin{equation}
 \frac{\mathrm{d}N}{\mathrm{d}X} = \int_{10^{20.3}}^\infty f(N, X) \mathrm{d} N\,,
\end{equation}
while $\Omega_\mathrm{DLA}$ is the integral of its first moment
\begin{equation}
 \Omega_\mathrm{DLA} = \frac{m_\mathrm{P} H_0}{c \rho_c}\int_{10^{20.3}}^\infty N f(N, X) \mathrm{d} N\,.
\end{equation}
Here $\rho_c$ is the critical density at $z=0$ and $m_\mathrm{P}$ is the proton mass. Note that this definition differs by a factor of $X_H = 0.76$, the 
primordial hydrogen mass fraction, from the quantity quoted by \cite{Noterdaeme:2012}. The difference comes from the fact that they divide the total mass
of neutral hydrogen by the hydrogen mass fraction in order to obtain the total gas mass in DLAs. However, this assumes 
the neutral fraction of hydrogen is unity, while in fact a significant fraction of the hydrogen in DLAs may be molecular, implying that total gas mass in DLAs 
is greater than $\Omega_\mathrm{DLA} / 0.76$ by an uncertain amount. We therefore quote the total HI mass.
Because the CDDF falls off sharply above the DLA threshold, $\frac{\mathrm{d}N}{\mathrm{d}X}$ is dominated by systems with $\NHI \sim 5\times 10^{20}$ \NHunit, while
$\Omega_\mathrm{DLA}$ has a stronger contribution from higher column densities. This explains why NOSN and DEF have similar values for $\frac{\mathrm{d}N}{\mathrm{d}X}$;
Figure \ref{fig:cddf_feed_rel} shows that their CDDFs cross at $\NHI \sim 5\times 10^{20}$ \NHunit. 
This is a coincidence, which would not be true for a different column density threshold.

Quoted errors in Figure \ref{fig:line_dens} are statistical. Systematic errors certainly dominate for $\frac{\mathrm{d}N}{\mathrm{d}X}$ 
(P. Noterdaeme, private communication), and are probably significant for $\Omega_\mathrm{DLA}$.\footnote{Note that identifying DLAs 
in low-resolution SDSS spectra is significantly more difficult than for the high resolution spectra used to measure the DLA metallicity.}
At $z=3$, Figure \ref{fig:line_dens} shows the trends from Figure \ref{fig:cddf_feed}. DEF, 2xUV and FAST are in good agreement with the data. HVEL predicts too 
little neutral hydrogen and too little DLA cross-section, while NOSN predicts too high a value for $\Omega_\mathrm{DLA}$. 

\begin{figure}
\includegraphics[width=0.45\textwidth]{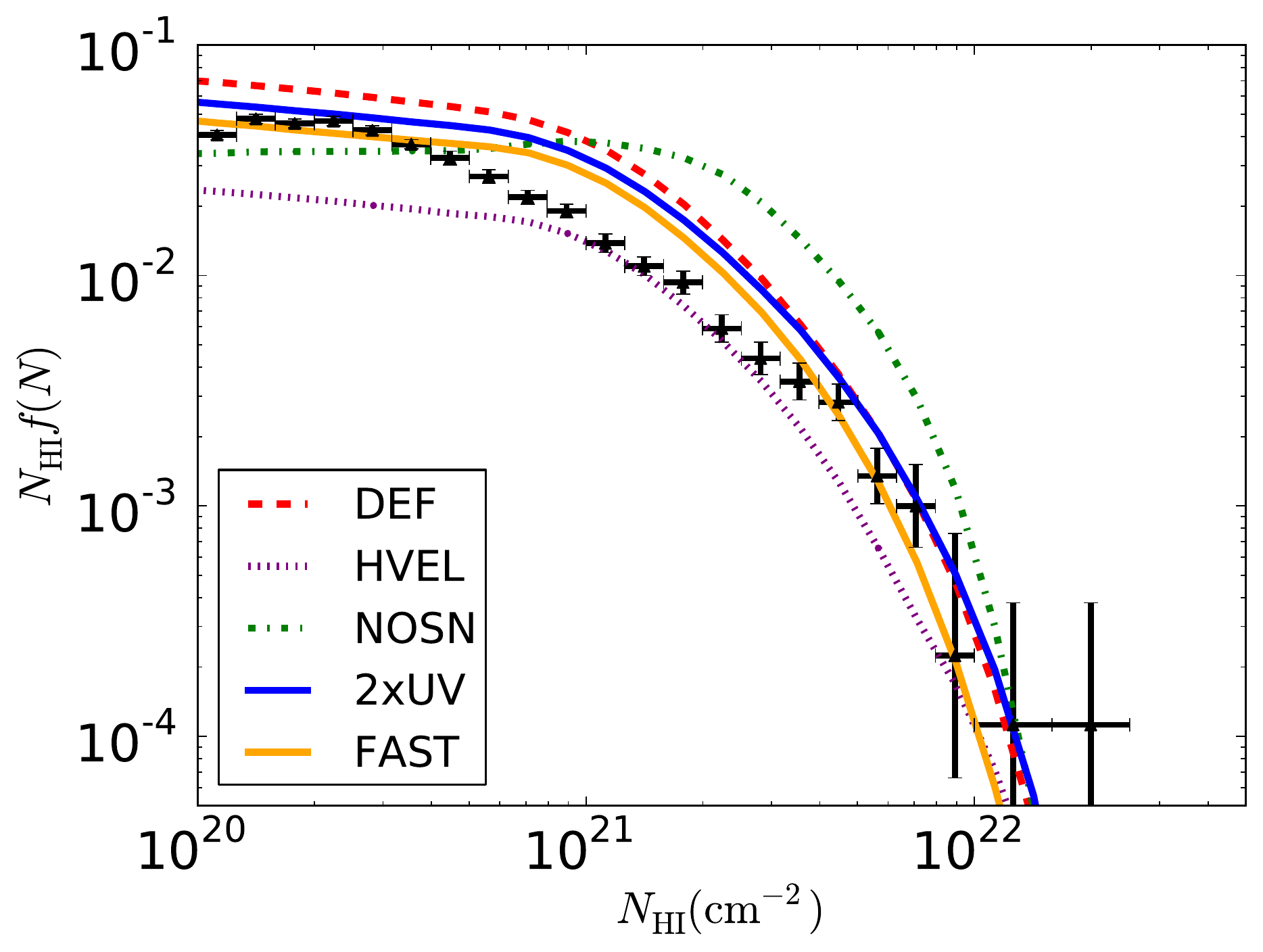}
\caption{The first moment of the CDDF at $z=2.5$, compared to observational data from \protect\cite{Noterdaeme:2012} (triangles), 
for our simulations. $\NHI$ is in physical cm$^{-2}$.}
\label{fig:cddf_z25}
\end{figure}

The three simulations with a large mass loading in small halos, 2xUV, DEF and FAST, all have very weak redshift evolution 
between $z=4$ and $z=3$, reproducing observational trends and reflecting a balance between gravitational collapse and supernova feedback.
However, between $z=2$ and $z=3$, both $\frac{\mathrm{d}N}{\mathrm{d}X}$ and $\Omega_\mathrm{DLA}$ begin to increase, a trend not seen in the observations.
Figure \ref{fig:cddf_z25} shows our results at $z=2.5$, compared to the CDDF from \cite{Noterdaeme:2012} at that redshift.
The excess HI is concentrated in systems with column densities between $5\times 10^{20}$ and $5\times 10^{21}$\NHunit. 
and thus is only weakly affected by the change in the photo-ionisation rate between DEF and 2xUV.
The FAST and HVEL simulations are closest to the observed $\Omega_\mathrm{HI}$ at $z=2.5$, each within $50$\% of the observed value.
Figure \ref{fig:cddf_z25} reveals that they do not match the observed shape of the CDDF at $z=2.5$, as they suppress its amplitude
relative to DEF, where what is required is an alteration of the shape to suppress a particular column density range. An interpolation between them 
could potentially match $\Omega_\mathrm{HI}$, but would produce too many columns with $\NHI \sim 10^{21}$\NHunit and too few with $\NHI \sim 10^{20}$ \NHunit.

\begin{figure}
\includegraphics[width=0.45\textwidth]{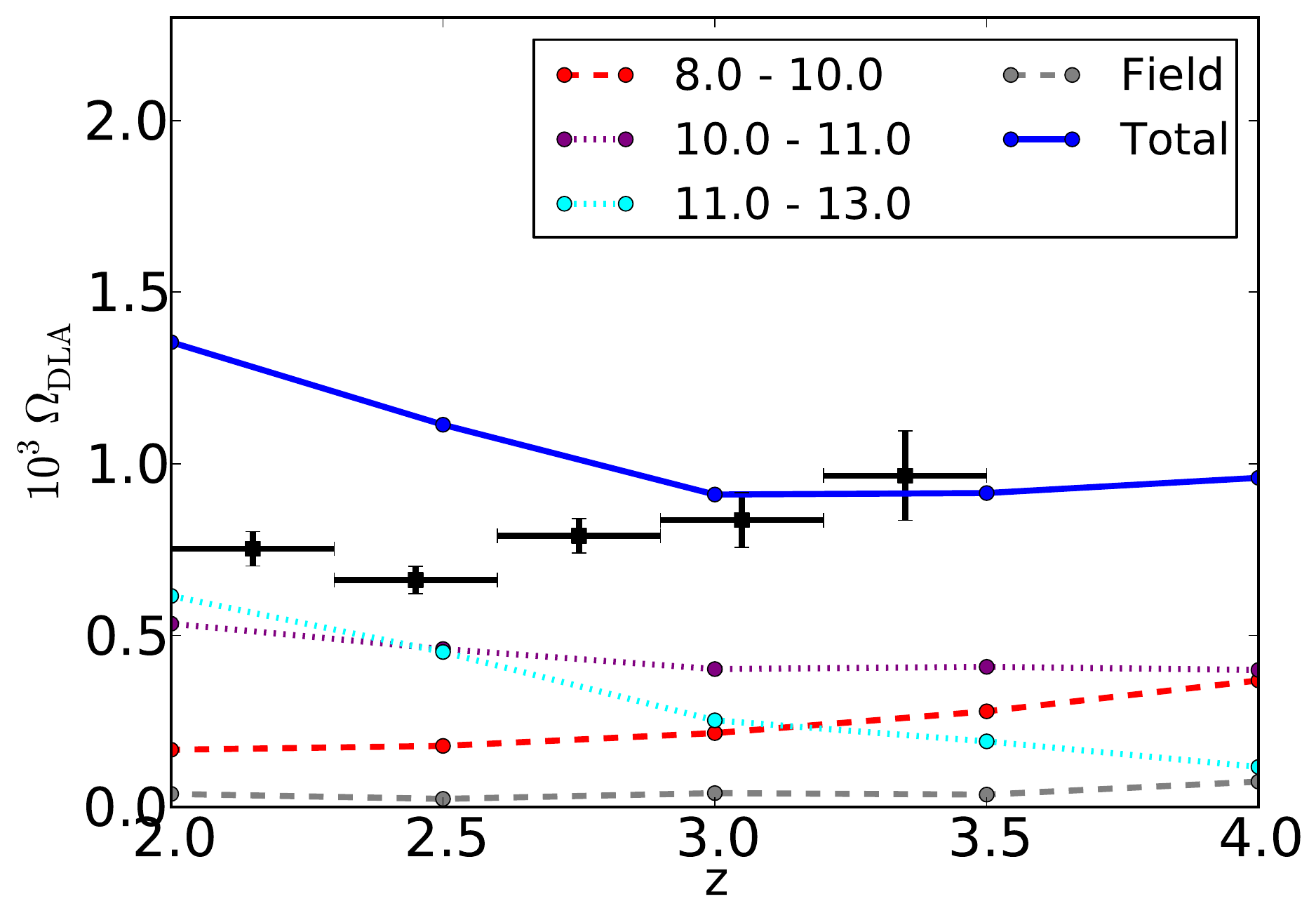}
\caption{Contributions to $\Omega_\mathrm{DLA}$ from different halo mass bins as a function of redshift for the 2xUV simulation.
The increase at $z=2.5$ is due to the formation of more massive halos. The field category denotes DLAs we could not associate with a halo.
Black squares show the observational data of \protect\cite{Noterdaeme:2012}.}
\label{fig:omega_hi_break}
\end{figure}

Figure \ref{fig:omega_hi_break} shows the contribution to $\Omega_\mathrm{DLA}$ from different halo masses, revealing that the increase at $z=2.5$ is coming 
predominantly from the formation of halos with $M > 10^{10} \Msun$. Thus a form of feedback able to 
suppress the amount of HI in these more massive halos is required. V13 found that their wind model over-produces the galaxy stellar mass function in halos 
of this size at $z=0$, compared to the observations of \cite{Baldry:2008}. There is thus also some evidence for stronger feedback in the 
galactic stellar component. Alternatively, the gas may be more ionised than we have assumed, perhaps by a thermal component to the feedback. 
We have checked that one cannot achieve agreement with observations by choosing a minimum wind velocity 
intermediate between HVEL and DEF. This suppresses DLAs in small halos, but has a reduced effect on the more massive halos 
which are responsible for the increase in $\Omega_\mathrm{DLA}$ at $z<3$. We also considered adding a feedback component 
in which the wind mass loading scales as the inverse of the wind velocity (momentum driven winds, following \cite{Dave:2013}), but found that a momentum driven component
large enough to affect $\Omega_\mathrm{DLA}$ completely suppressed columns with $\NHI > 10^{21}$ \NHunit, while leaving lower columns unchanged, and 
so was strongly discrepant with the observed column density function.

\subsection{DLA Bias}
\label{sec:dlabiasres}

\begin{figure}
\includegraphics[width=0.45\textwidth]{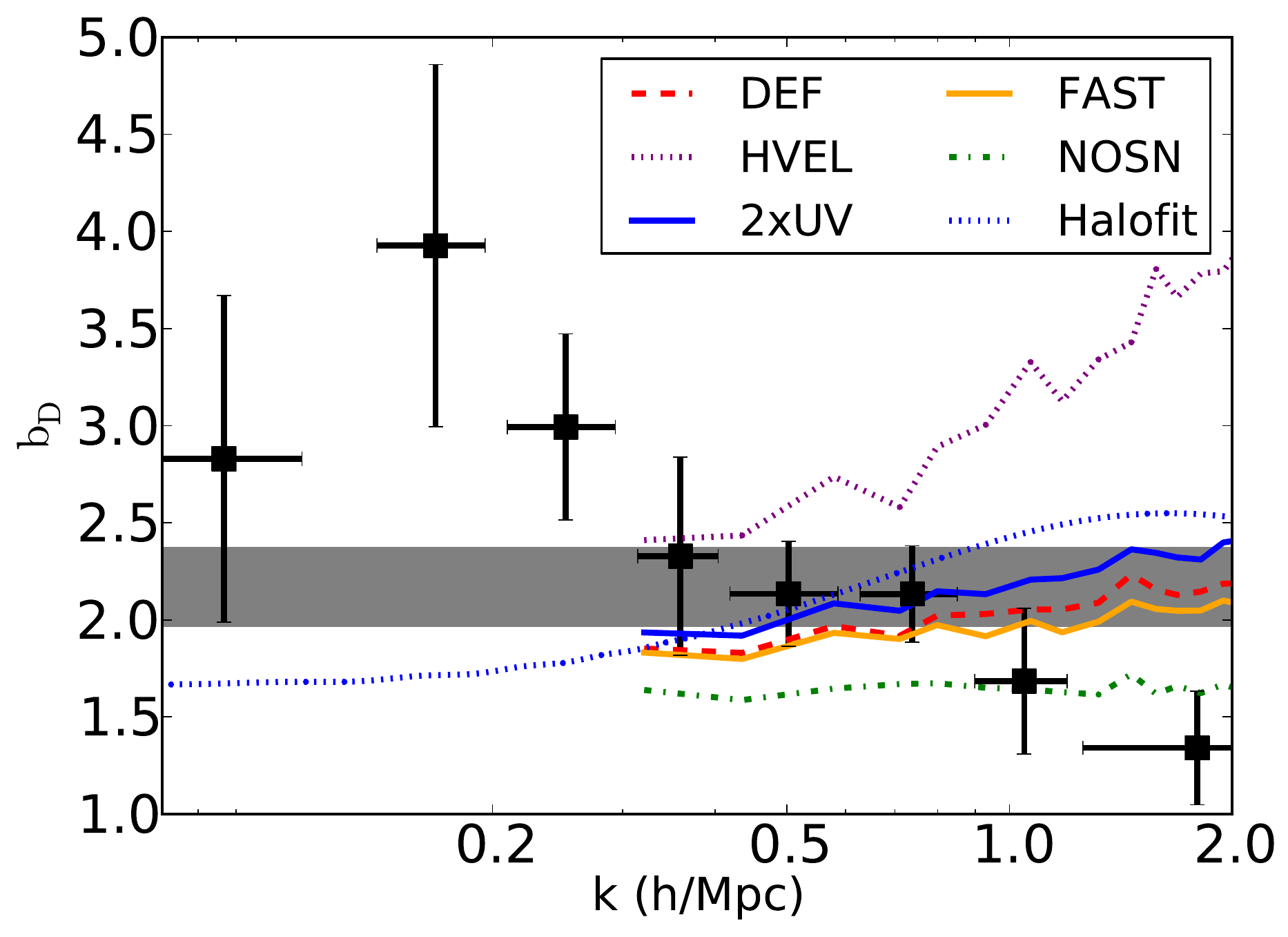}
\caption{The bias of the DLA power spectrum to the matter power spectrum at $z=2.3$, as measured from our simulations.
Points with error bars show the bias measured by \protect\cite{FontRibera:2012} as a function of scale, while 
the grey band shows a $1\,\sigma$ confidence interval for the average bias on scales $k < 1.2 \hMpc$.
The dotted line labelled Halofit shows the results of applying Eq. (\ref{eq:dlanlbias}) to the 2xUV simulation.}
\label{fig:dla_bias}
\end{figure}

We define the DLA bias in Eq. (\ref{eq:dlabias}) as the ratio of the DLA power spectrum to the non-linear matter power spectrum.
Figure \ref{fig:dla_bias} shows the results of our simulations and the measurements of \cite{FontRibera:2012} (henceforth F12). 
The 2xUV simulation is in good agreement with the observations on the scales where the two overlap.
The DLA bias in DEF and FAST is a little lower than is observed, as both simulations are less efficienct than 2xUV at suppressing the formation of DLAs in low-mass halos.
The HVEL simulation yields a strongly scale-dependent DLA bias, which comes close to the observations on scales of the simulation box, but is significantly larger on smaller scales.
Recently, \cite{Kuhlen:2012, Kuhlen:2013} proposed a model without stellar feedback, where star formation is instead suppressed by 
raising the star formation threshold density. Such a model would produce DLAs with a distribution similar to our NOSN simulation, and thus
have a DLA bias significantly smaller than is observed.

F12 used their measurements to deduce the mass of DLA host halos. To do this, they assumed scale independence of the DLA bias in the large-scale limit, a
halo mass function and a model for the abundance of DLAs in a halo. The DLA bias is in this linear model is then
\begin{equation}
 b_\mathrm{DLA} = \frac{\int b(M) \sigma_\mathrm{T} (M) dM }{\int \sigma_\mathrm{T} (M) dM } \,,
\label{eq:dlalinearbias}
\end{equation}
where $b(M)$ is the linear halo bias \citep{Tinker:2010}, and $\sigma_\mathrm{T}$ is the total DLA cross-section 
in a halo mass bin. Note that $\sigma_\mathrm{T} (M) = n(M) \sigmaDLA(M)$, where $n(M)$ is the halo mass function\citep{Tinker:2008},
and $\sigmaDLA$ the DLA cross-section for a given halo mass.
F12 described the DLA cross-section using the results of \cite{Pontzen:2008}, which has $\sigmaDLA \sim M^{0.5}$ for $M > 10^{10} \Msun$.
As discussed in Section~\ref{sec:halosize}, our 2xUV simulation has $\sigmaDLA \propto M$, which yields a significantly increased DLA cross-section in 
more massive halos, and thus a larger DLA bias.

Applying Eq.~(\ref{eq:dlalinearbias}) to 2xUV gives $b_\mathrm{DLA} = 1.72$ at $z=2.3$, about $15\%$ less than the mean bias shown in 
Figure~\ref{fig:dla_bias}. This discrepancy arises due to non-linear growth in the DLA power spectrum.
For our cosmology at $z=2.3$, the non-linear scale (defined as the scale where the variance of the fluctuations is unity)
is $k \approx 2.4 \hMpc$. However, for objects which trace the matter power spectrum with a bias of two, the non-linear scale is four times larger, $k \sim 0.6 \Mpch$, 
and non-linear growth can affect the power spectrum at the $10\%$ level for $k \sim 0.2 \Mpch$.

As the rate of structure growth is higher in the non-linear regime, non-linear growth leads to an increase in bias on small scales.
Since more massive halos are more biased and thus more non-linear, non-linear growth increases their fractional contribution to the DLA bias.
We shall account for these effects by making the (crude) assumption that the non-linear power spectrum of a biased object, $P_\mathrm{NL}$,
can be described by applying the {\small HALOFIT} model, $\mathcal{H}$, \citep{Smith:2003} to the biased linear power spectrum.
That is, we assume $P_\mathrm{NL} = \mathcal{H}\left[b^2 P(k)\right]$. 
The modified form of Eq. (\ref{eq:dlalinearbias}) is then
\begin{equation}
 b^2_\mathrm{DLA}(k) = \frac{\int \mathcal{H}\left[ b^2(M)P(k) \right] \sigma^2_\mathrm{T} dM }{\int \mathcal{H}\left[ P(k) \right]  \sigma^2_\mathrm{T} dM } \,.
\label{eq:dlanlbias}
\end{equation}
Figure~\ref{fig:dla_bias} shows Eq.~(\ref{eq:dlanlbias}) applied to the 2xUV simulation. While this model overestimates non-linear effects on small scales, 
it can reconcile Eq.~(\ref{eq:dlalinearbias}) with the direct calculation.
It does not, however, agree with the scale-dependence of the observed bias, which decreases for $k > 0.8 \hMpc$. This may be due to the effects of noise, or 
to differences between the DLA power spectrum and the cross-correlation with the \Lya forest. Computing the cross-correlation 
directly is beyond the scope of this paper, but we plan to examine it further in future work. For now, we note only that the physical mechanism causing 
this effect may conceivably impact larger scales and thus change our agreement with the observed data.

Since the good agreement of the 2xUV simulation with observations results in part from non-linear growth,
we expect the bias to decrease on larger scales, and asymptote to the linear theory value by $k = 0.1 \hMpc$.
This will reduce the mean bias derived from the simulations; a simple estimate of the effect using Eq.~(\ref{eq:dlanlbias}) suggests that the simulaton 
would be around $1.5 \sigma$ below the observations. However, since most of the statistical power for the F12 measurement is derived
from scales probed by our simulation box, the discrepancy in a full likelihood analysis would probably be somewhat less.

\subsection{Metallicity}
\label{sec:metallicity}

\begin{figure*}
\includegraphics[width=0.33\textwidth]{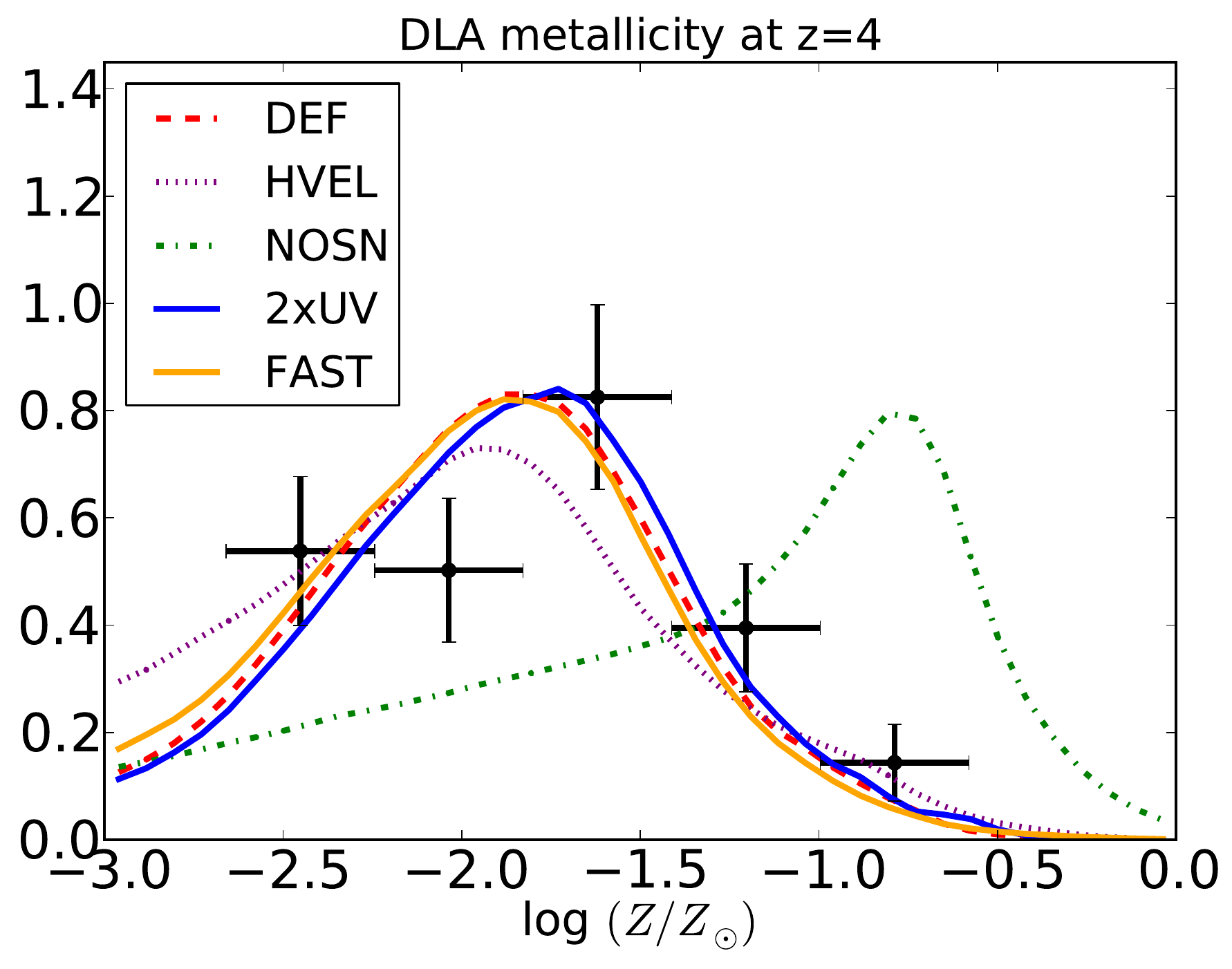}
\includegraphics[width=0.33\textwidth]{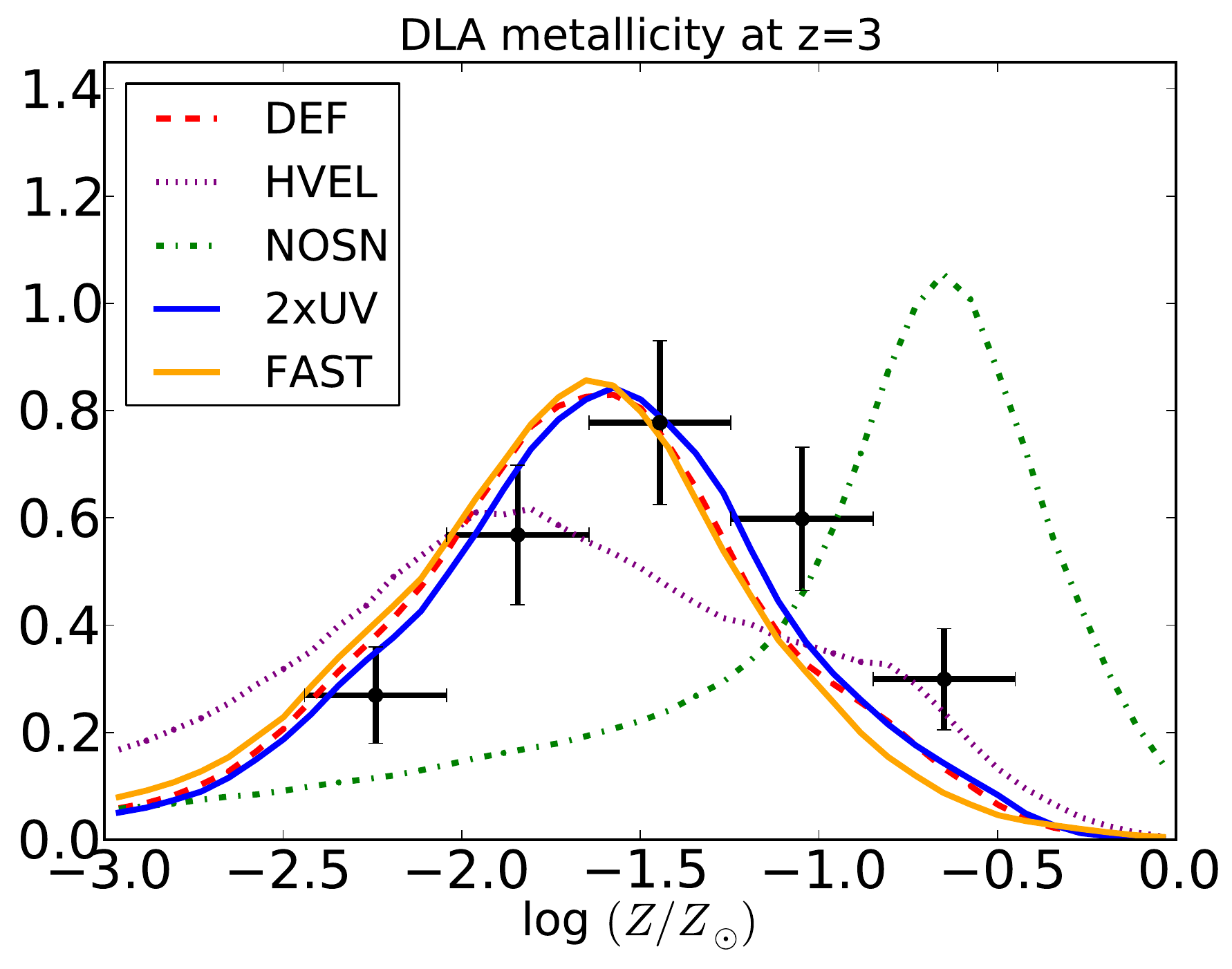}
\includegraphics[width=0.33\textwidth]{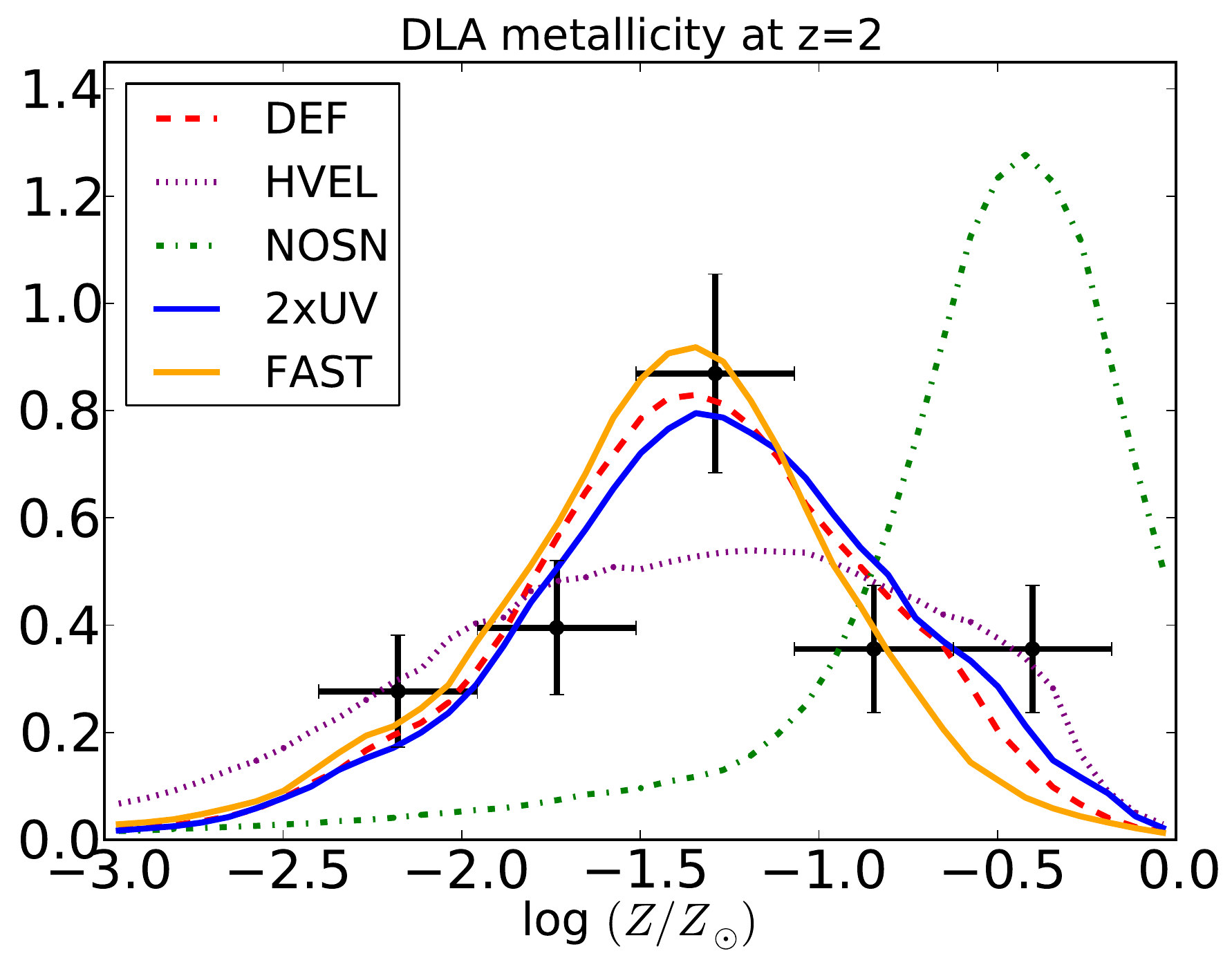}
\caption{The probability density function of the log of DLA metallicity normalised so that the total area under the curve is unity. 
(Left) $z=4$ (Centre) $z=3$ (Right) $z=2$. Data is from the compendium found in \protect\cite{Rafelski:2012}. $Z_\odot = 0.0182 = 0.0134/0.7381$. }
\label{fig:metals}
\end{figure*}

Figure \ref{fig:metals} shows the probability density function of the log of mass-weighted DLA metallicity, normalised 
so that the total area under the curve is unity. We define the metallicity of each DLA grid cell to be
\begin{equation}
 Z = \frac{\sum M_{Z}}{\sum M_{H}}\,,
\end{equation}
where $M_{Z}$ is the mass in metals of a pixel and $M_{H}$ the total gas mass.
All metallicities are quoted in units of the Solar metallicity, which we take to be $M_{Z_\odot} / M_{H_\odot} = 0.0134/0.7381$ \citep{Asplund:2009}.
\footnote{Note that in \cite{Vogelsberger:2014} we incorrectly used a Solar metallicity of $0.0134$.}
The total mass in metals and the total gas mass in each cell are computed separately using the same method used to compute HI columns and
explained in Section \ref{sec:colden}. 

DEF, 2xUV and FAST give very similar results in very good agreement with the data at all three redshifts.
NOSN is a very poor fit to the data, giving a mean DLA metallicity much 
higher than observed. While HVEL is in good agreement at $z=4$, 
at $z=3$ it does not reproduce the peak in the metallicity, instead yielding a plateau.
Figure \ref{fig:dla_halos} shows that this plateau appears also in the distribution of DLA host halos,
and we have checked that the differences in metallicity between HVEL and DEF are dominated by changes in the DLA host halos,
rather than variations in the DLA mass-metallicity relation.

The mean metallicity evolves with redshift, increasing at later times. 
This mirrors evolution in the host halos of DLAs shown in Figure \ref{fig:dla_halohist}. At later times DLAs trace more massive halos, which are more metal-rich. 
The DLA metallicity is well described by a power law in halo mass, whose parameters show little evolution with redshift,
confirming that the evolution in the average DLA metallicity is driven by evolution in the host halo distribution.

We have checked that the metal loading factor of the winds does not affect the DLA metallicity, even at redshift four, 
where re-accretion will not be significant. Thus DLAs are primarily enriched directly by the stellar mass in the host halo, 
rather than through outflows. Our good agreement with the observed DLA metallicity implies that our DLA halo hosts 
have the correct stellar mass. Note however that these DLAs are hosted in small halos, so their stellar components are hard to observe directly.

\section{Conclusions}
\label{sec:conclusions}

We have studied Damped \Lya systems (DLAs) using a suite of hydrodynamic simulations which vary the parameters of the stellar feedback model 
around those used in \cite{Vogelsberger:2013} (V13). The predictions of this model for DLAs have not previously been examined, 
and so our work serves as an important verification of its ability to match observed data. By using a suite of simulations, 
we are able to show the extent to which each observation constrains our model, something which has only previously been done for the column density function. 

Turning to the DLA metallicity distribution, most of our simulations gave similar results, in good agreement with the observations at $z=2-4$.
This suggests that the DLA metallicity is not particularly sensitive to the detailed parameters of the feedback model.
The exceptions were a simulation without feedback and one including outflows with a constant velocity in small halos.
Similar models tend to over-produce the faint end of the galaxy stellar mass function at low redshift~\citep{Puchwein:2013}.
Since the metallicity distribution traces the stellar mass of DLA hosts, it provides a valuable indirect probe of the 
faint end of the stellar mass function at redshifts where it cannot yet be observed directly.
That the two simulations mentioned above do not match the DLA metallicity therefore demonstrates the need, even at $z=2-4$, 
for a form of feedback which suppresses star formation very efficiently in small halos. 

The observed abundance of neutral hydrogen in DLAs is more constraining.
The parameters chosen in V13 over-produce the number of weak DLAs and LLS relative to observations at $z=3$.
We demonstrate that by increasing the photo-ionisation rate from the UV background to match constraints from improved measurements 
of the IGM temperature \citep{Becker:2013}, we are able to alleviate the problem, both directly 
by ionising the gas and indirectly by preventing the accretion of small halos.
With this modification, we match the observed column density function at $z=3$ and its integrals,
the total density of neutral hydrogen and the incident rate of DLAs, at $z=3-4$.

However, for $z < 3$ the total HI density in our simulations increases, and by $z=2.5$ all our simulations 
are in strong tension with observations. By comparing to the column density function (CDDF) at $z=2.5$, we demonstrate 
that this tension arises from growth in the amount of HI in halos of mass $10^{11} - 10^{13} \Msun$. 
We considered a minimum wind velocity of $600$ \kms, which makes the wind velocity constant in most DLA host halos.
However, we found that, although we now reproduce the observed evolution in $\Omega_\mathrm{DLA}$ at $z > 2$, the stronger winds over-suppress 
the DLA cross-section, producing a normalisation of $\Omega_\mathrm{DLA}$ which is too low. 
We considered other variations, including a simulation with a $50$\% higher wind velocity. This simulation does produce somewhat better agreement with $\Omega_\mathrm{DLA}$ 
at $z=2.5$, but it does not match the detailed shape of the CDDF, producing too few weak DLAs and too many stronger ones. 
Resolving this tension could involve winds which behave like our model in smaller halos, but are 
effective at reducing the density of gas in more massive halos. Alternatively, the feedback could become yet stronger in small halos at low redshift.

We compare recent measurements of the DLA bias from the cross-correlation of DLAs with the \Lya forest to our simulations.
Due to our strong feedback, the DLA cross-section in our preferred models is dominated by systems hosted
in halos of mass $10^{10} - 10^{11}\Msun$, and we find that our preferred model produces simulations and observations in good agreement on scales where they overlap. 
However, a simple linear bias model applied to our simulation suggests a bias around $2-\sigma$ lower than observed.
We show that this discrepancy is due to non-linear effects, which on the scales concerned can increase the DLA bias by $15\%$.
Structure formation is faster in the non-linear regime; hence the larger and more massive halos make a larger contribution to the total power spectrum, 
increasing the DLA bias. Non-linear growth should therefore be taken into account when interpreting observational results.

The  bias measurements, unlike our simulated results, appear to decrease substantially for $k > 1 \hMpc$. 
The reason for this is currently unclear, but may be related to the fact that we have derived the 
DLA bias directly from the DLA power spectrum, while the observed bias is derived from the cross-correlation of DLAs with the \Lya forest.

We have deferred discussion of the velocity structure of DLAs, which provide an important extra constraint on the model, 
to a companion paper \citep{Bird:2014a}.  There we will show that we are able to use the same models which are preferred in 
this paper to match the velocity width distribution significantly better than previous work, corroborating the need for strong 
feedback in low mass haloes at $z=2-4$.

To summarise, our detailed comparison of observed DLA properties with hydrodynamic simulations has shown 
that the observed incidence rate and bias provide particularly powerful constraints on the parameters of the stellar
feedback model in the simulations. The default feedback model presented in V13 (with the UVB amplitude doubled to match the most recent UVB measurements) 
produces DLAs in good agreement with a wide range of observed DLA properties. 
Some discrepancies remain, particularly at $z < 3$, but overall our results confirm  a picture where DLAs form in a wide range 
of halos with a relatively modest mean mass, $10^{10} - 10^{11}\Msun$. We have shown explicitly that this mass range
is not in disagreement with the recently reported value of the DLA bias parameter, despite the indications of simple linear calculations, 
due to non-linear growth of the biased DLA host haloes in the simulations.

\section*{Acknowledgements}

SB thanks Pasquier Noterdaeme for useful discussions regarding the data, 
J.~Xavier Prochaska for useful comments and proof-reading an earlier version of the paper
and Ryan Cooke for noticing a units error in an earlier draft.
SB is supported by the National Science Foundation grant number AST-0907969, 
the W.M. Keck Foundation and the Institute for Advanced Study.
MGH acknowledges support from the FP7 ERC Advanced Grant Emergence-320596. 
VS acknowledges support from the European Research Council under
ERC-StG grant EXAGAL-308037. LH is supported by NASA ATP Award NNX12AC67G
and NSF grant AST-1312095.

\label{lastpage}
\bibliography{DLAfeedback}

\end{document}